\documentclass[review]{elsarticle}

\usepackage{lineno,hyperref}
\usepackage{graphicx}
\usepackage{subfig}
\usepackage[linesnumbered,ruled,lined]{algorithm2e}
\usepackage{float}
\usepackage{tikz}
\usepackage{epstopdf}
\usepackage{amssymb}
\usepackage{amsmath}
\usetikzlibrary{arrows.meta}

\newcommand{\tr}{\mathop{\mathrm{tr}}}

\journal{J. X. X}

\begin{document}

\begin{frontmatter}

\title{A multi-resolution SPH method for fluid-structure interactions}

\author{Chi Zhang}
\ead{c.zhang@tum.de}
\author{Massoud Rezavand}
\ead{massoud.rezavand@tum.de}
\author{Xiangyu Hu \corref{mycorrespondingauthor}}
\cortext[mycorrespondingauthor]{Corresponding author. Tel.: +49 89 289 16152.}
\ead{xiangyu.hu@tum.de}
\address{Department of Mechanical Engineering, 
	Technical University of Munich, 85748 Garching, Germany}

\begin{abstract}
In this paper, 
we present a multi-resolution smoothed particle hydrodynamics (SPH) method for modeling fluid-structure interaction (FSI) problems. 
By introducing different smoothing lengths and time steps,
the spatio-temporal discretization is applied with different resolutions for fluid and structure. 
To ensure momentum conservation at the fluid-structure coupling, 
a position-based Verlet time integration scheme is introduced. 
Furthermore, 
the time-averaged velocity and acceleration of solid particles are introduced to enhance force matching in the fluid and solid equations. 
A set of numerical examples including several bio-mechanical problems are considered to demonstrate the efficiency, 
accuracy and robustness of the present method.  
A open-source code for all the examples is also provided.
\end{abstract}

\end{frontmatter}

%
%
\clearpage
\section{Introduction}
Fluid-structure interaction (FSI) problems, 
in particular those involving flexible and deformable elastic structures, 
are ubiquitous in natural phenomena, e.g. 
aerial animal flying and aquatic animal swimming, 
and also play a key role in a vast range of engineering examples, e.g. 
hydroelastic slamming, energy harvesting and prosthetic heart valves. 
Due to the intrinsic complexity of the surrounding flow and the large deformation of the structure, 
computational study of FSI problems is highly challenging. 

Mesh-based approaches, 
such as finite element method (FEM) \cite{tezduyar1992new} 
and immersed-boundary method (IBM) \cite{peskin2002immersed}, 
can encounter difficulties for simulating FSI problems. 
Typical difficulties for FEM method include the treatment of the convective terms 
and the updating of the meshes for moving fluid-structure interface especially when large structure deformation is taken into account \cite{onate1996finite}. 
IBM utilizes two overlapped Lagrangian and Eulerian meshes. 
While the fluid equation is solved on the Eulerian mesh, 
the effects of solid structure are taken into account by distributing the forces computed on the deformed Lagrangian mesh to the Eulerian mesh using a kernel function. 
Such fairly weak coupling formulation leads to the Lagrangian-Eulerian mismatches on the kinematics and the distribution of solid structure forces.

An alternative approach for tackling FSI problems is to discretize both the fluid and solid equations by using Lagrangian meshfree methods, 
for instance smoothed particle hydrodynamics (SPH) \cite{lucy1977numerical, gingold1977smoothed, zhang2017weakly}, 
moving-particle semi-implicit method (MPS) \cite{koshizuka1996moving} and discrete element method (DEM) \cite{mishra1992discrete}.
Owing to their peculiar advantages in handling material interfaces \cite{zhang2017generalized, rezavand2019weakly}, 
and the capability of capturing violent events such as impact and breaking \cite{zhang2019weakly}, 
Lagrangian particle-based methods can be conspicuously effective when studying FSI problems, 
especially when they are characterized by large displacement of fluid-structure interfaces.
In recent years, 
promising results have been obtained for the simulation of FSI problems with particle-based methods 
\cite{antoci2007numerical, oger2009simulations, liu2013numerical, rafiee2009sph, ren2013sph, wang2019sph, zhan2019stabilized, han2018sph}, 
or combined with FEM method \cite{yang2012free, zhang2019mps, chen2019numerical}.

When the SPH method is used for studying FSI problems, 
single spatio-temporal resolution is commonly employed for the discretization of fluid and solid equations \cite{antoci2007numerical}. 
Specifically, 
a uniform particle spacing is used for the entire computational domain and a single time step, 
which is the smallest of these required by the fluid and solid structure \cite{antoci2007numerical}, 
is applied for time integration. 
From the computational efficiency point of view, 
the single-resolution approach can be rather expensive in applications 
where the structure requires small time steps. 
This is the case when the structure is relative thin, has fine resolutions or has a high Poisson ratio.
Therefore, 
developing a multi-resolution SPH method capable of employing various spatio-temporal resolutions for fluid and solid structure is desirable.

In this paper, we propose a multi-resolution SPH method to increase the computational efficiency for simulating FSI problems. 
Following the unified SPH discretization in Ref. \cite{han2018sph}, 
the update Lagrangian formulation is utilized for the fluid dynamics, 
and the total Lagrangian formulation for solid dynamics. 
Since the solid structure is able to be resolved at a higher spatial resolution, 
the computational efficiency is enhanced when a lower resolution discretization for the fluid is sufficient. 
To further improve the efficiency, 
different time steps are utilized for integrating the fluid and solid equations.  
A position-based Verlet time integration scheme is proposed for enforcing the momentum conservation at the fluid-structure coupling. 
The velocity and acceleration of solid particles are averaged over a fluid time step to enhance the force matching, 
e.g. calculates a set of reliable interaction forces over a single fluid time step during which structure state has been updated several times,  
in fluid and solid equations. 
Different with previous researches which attempted to apply different resolutions for the discretization of fluid equations 
\cite{owen1998adaptive, barcarolo2014adaptive, lastiwka2005adaptive, vacondio2013variable,hu2019consistent, omidvar2012wave, bian2015multi}, 
the objective of present method is to develop a multi-resolution SPH method for FSI problems. 
The efficiency, accuracy and robustness of the present method are validated with two benchmark tests, 
i.e.  flow-induced vibration and dam-break flow with elastic gate. 
Then, two bio-mechanical applications, i.e.
venous valve and fish flapping, 
are considered to demonstrate the versatility and potential of the present method in simulating bio-mechanical systems.
The remainder of this paper is arranged as follows: 
Section \ref{sec:governingeqs} briefly summarize the governing equations and preliminary SPH algorithm for simulating FSI problems. 
The multi-resolution method is detailed in Section \ref{sec:method}. 
Numerical validations and applications are presented and discussed in Section \ref{sec:validation}.
Concluding remarks are given in Section \ref{sec:conclusion} and all code and data-sets accompanying this work 
are available on GitHub at \url{https://github.com/Xiangyu-Hu/SPHinXsys}.
%
%
\section{Governing equations and preliminary work}  \label{sec:governingeqs}
%
Before moving to detailed description of the multi-resolution method, 
we first introduce the governing equations for fluid and solid dynamics, 
and then briefly summarize the corresponding discritzations with unified SPH formulation and more details are referred to Han and Hu \cite{han2018sph}.
\subsection{Fluid- and solid-dynamics equations}  
The governing equations in updated Lagrangian formulation for an isothermal and Newtonian fluid flow  are the mass conservation equation
\begin{equation}\label{mass-conservation-f}
\frac{\text{d} \rho}{\text{d} t}  = -\rho \nabla \cdot\mathbf{v} ,
\end{equation}
and the momentum conservation equation
\begin{equation}\label{momentum-conservation-f}
\rho \frac{\text{d} \mathbf{v}}{\text{d} t}  = -\nabla p + \eta \nabla^{2} \mathbf{v} + \mathbf{g},
\end{equation}
where $\rho$ is the density, 
$t$ the time,
$\mathbf{v}$ the velocity,
$p$ the pressure, 
$\eta$ the dynamics viscosity and $\mathbf{g}$ the body force.
To close the system of Eqs. \eqref{mass-conservation-f} and \eqref{momentum-conservation-f}, 
an appropriate equation of state (EoS) for the fluid is required.
Following the weakly-compressible assumption \cite{monaghan1994simulating,morris1997modeling} 
for modeling incompressible flow, 
the pressure is evaluated from the density through the Tait equation
\begin{equation} \label{eqeos}
p = \frac{\rho^0 (c^F)^2}{\gamma}\left( (\frac{\rho}{\rho^0})^{\gamma} - 1\right) ,
\end{equation}
where $\gamma = 7 $ and the reference density $\rho^0$. 
Here, $c^F$ is the artificial sound speed of fluid.  
Numerical experiments confirm that the density variations are limited to about $1 \% $ \cite{morris1997modeling} 
when $ c^F = 10 v_{max}$ with $v_{max}$ denoting the maximum anticipated flow speed. 
 
For the solid structure, an elastic and weakly-compressible material is taken into account. 
Following the convention in continuum mechanics, 
the initial position $\mathbf{r}^0$ of a material point is defined in the initial reference configuration, 
and the current position $\mathbf{r}$ in the deformed current configuration. 
Then the displacement $\mathbf{u}$ of a material point, 
which is obtained by the difference between its current position and initial reference position, 
can be obtained as
\begin{equation}\label{trajectory}
\mathbf{u} =  \mathbf{r} - \mathbf{r}^0 .
\end{equation}
Thus, the deformation tensor, is defined by 
\begin{equation}\label{deformation-tensor}
\mathbb{F}  = \nabla^{0} \mathbf{u}  + \mathbb{I} ,
\end{equation}
where $\mathbb{I}$ denotes the identity matrix, 
and $\nabla^{0} \equiv \frac{\partial}{\partial \mathbf{r}^0}$ stands for the gradient operator with respect to the initial reference configuration. 
In the total Lagrangian formulation, the mass conservation equation is given by
\begin{equation}\label{mass-conservation-s}
\rho  -  \rho^0 J ^{-1} = 0 ,
\end{equation}
where $\rho^0$ is the initial reference density and $J = \det(\mathbb{F})$ the Jacobian determinant of deformation tensor $\mathbb{F}$. 
The momentum conservation equation reads
\begin{equation}\label{momentum-conservation-s}
\rho^0 \frac{d \mathbf{v}}{d t}  =  \nabla^{0} \cdot \mathbb{P},
\end{equation}
where $\mathbb{P}$ denotes the first Piola-Kirchhoff stress tensor, 
which relates the forces in the current configuration to the areas in the initial reference configuration.
For an ideal elastic or Kirchhoff material, $\mathbb{P}$ is given by
\begin{equation}\label{linear-elasticity}
\mathbb{P} = \mathbb{F} \mathbb{S}, 
\end{equation}
where $\mathbb{S}$ represents the second Piola-Kirchhoff stress which is evaluated via the constitutive equation relating $\mathbb{F}$ with 
the Green-Lagrangian strain tensor $\mathbb{E}$ defined as 
\begin{equation}\label{Lagrangian-strain}
\mathbb{E} = \frac{1}{2} \left( \mathbb{F}^{T}\mathbb{F} - \mathbb{I}\right) .
\end{equation}

In particular, when the material is linear elastic and isotropic, the constitutive equation is simply given by
\begin{eqnarray}\label{isotropic-linear-elasticity}
\mathbb{S} & = & K \tr\left(\mathbb{E}\right)  \mathbb{I} + 2 G \left(\mathbb{E} - \frac{1}{3}\tr\left(\mathbb{E}\right)  \mathbb{I} \right) \nonumber \\
& = & \lambda \tr\left(\mathbb{E}\right) \mathbb{I} + 2 \mu \mathbb{E} ,
\end{eqnarray}
where $\lambda$ and $\mu$ are Lem$\acute{e}$ parameters, 
$K = \lambda + (2\mu/3)$ the bulk modulus and $G = \mu$ the shear modulus. 
The relation between the two modulus is given by
\begin{equation}\label{relation-modulus}
E = 2G \left(1+2\nu\right) = 3K\left(1 - 2\nu\right)
\end{equation}
with $E$ denotes the Young's modulus and $\nu$ the Poisson ratio. 
Note that the sound speed of solid structure is defined as $c^{S} = \sqrt{\frac{K}{\rho}}$. 
In present work, 
a Neo-Hookean model defined by the strain-energy density function\cite{ogden1997non}
\begin{eqnarray}\label{Neo-Hookean-energy}
W  =  \mu \tr \left(\mathbb{E}\right) - \mu \ln J + \frac{\lambda}{2}(\ln J)^{2}
\end{eqnarray}
is used for predicting the nonlinear stress-strain behavior of materials undergoing large deformations. 
Note that the second Piola-Kirchhoff stress $\mathbb{S}$ can be derived as 
\begin{equation}\label{2rd-PK}
\mathbb{S} = \frac{\partial W}{\partial \mathbb{E}}
\end{equation}
from strain-energy density function.  
\subsection{Preliminary work}
A SPH discretization of the mass and momentum conservation equations, 
Eqs. \eqref{mass-conservation-f}  and 
\eqref{momentum-conservation-f}, 
can be respectively derived as
\begin{equation} 
\frac{\text{d} \rho_i}{\text{d} t}  = 2\rho_i \sum_{j}  V_j (\mathbf{v}_i - \widetilde{\mathbf{v}}_{ij}) \nabla_i W_{ij},
\end{equation}
and
\begin{equation}\label{fluid-model}
m_i \frac{\text{d} \mathbf{v}_i}{\text{d} t}  =  - 2  \sum_j V_i V_j \widetilde{p}_{ij} \nabla_i W_{ij} + 
2\sum_j \eta V_i V_j \frac{{{\mathbf{v}}_{ij}}}{{{r}_{ij}}} \frac{\partial {{W}_{ij}}}{\partial {{r}_{ij}}} + \mathbf{g} +
\mathbf{f}_i^{S:p} + \mathbf{f}_i^{S:v},
\end{equation}
where $V_i$ denotes the particle volume and 
$\nabla_i W_{ij} = \nabla_i W(\mathbf{r}_{ij}, h) = \frac{\mathbf{e}_{ij}}{r_{ij}} \frac{\partial {{W}_{ij}}}{\partial {{r}_{ij}}} $, 
where $\mathbf{r}_{ij} = \mathbf{r}_i - \mathbf{r}_j$ and $h$ is smoothing length, 
the unit vector $\mathbf{e}_{ij} = \frac{\mathbf{r}_{ij}}{r_{ij}}$, 
representing the derivative of the kernel function with respect to $\mathbf{r}_i$, 
the position of particle $i$.
Here, the inter-particle velocity and pressure averages, 
$\widetilde{\mathbf{v}}_{ij}$ and $\widetilde{p}_{ij}$, 
can be obtained by a simple average, 
or from the solution of the inter-particle Riemann problem. 
In the present work, 
a low-dissipation Riemann solver is applied and we refer to Ref. \cite{zhang2017weakly} for more details.
The $\mathbf{f}_i^{S:p}$ and $\mathbf{f}_i^{S:v}$ terms are the pressure and viscous forces 
acting on the fluid particle, respectively, due to the presence of the solid structure.  

To discretize the governing equations of solid dynamics, 
a correction matrix $\mathbb{B}^0$ \cite{vignjevic2006sph, han2018sph} is first introduced as
\begin{equation} \label{correctmatrix}
\mathbb{B}^0_a = \left( \sum_b \left( \mathbf{r}_b^0 - \mathbf{r}_a^0 \right) \otimes \nabla_a^0 W_{ab} \right) ^{-1} ,
\end{equation}
where 
\begin{equation}\label{strongkernel}
\nabla_a^0 W_{a} = \frac{\partial W\left( |\mathbf{r}_{ab}^0|, h \right)}  {\partial |\mathbf{r}_{ab}^0|} \mathbf{e}_{ab}^0
\end{equation}
stands for the gradient of the kernel function evaluated at the initial reference configuration. 
Note that subscripts $a$ and $b$ are used to denote solid particles, 
and smoothing length $h$ is identical to that in the discretization of fluid equations of Eq. \eqref{fluid-model}.   
Using the initial configuration as the reference, 
the solid density is directly evaluated as 
\begin{equation}\label{solid-con-sph}
\rho_a = \rho_a^0 \frac{1}{J}
\end{equation}
from Eq. \eqref{mass-conservation-s}.
We can now discretize Eq. \eqref{momentum-conservation-s} in the following form 
\begin{equation}\label{mom-sph}
m_a \frac{\text{d} \mathbf{v}}{\text{d}t} = 2 \sum_b V_a V_b \tilde{\mathbb{P}}_{ab} \nabla_a^0 W_{ab} +\mathbf{g} + \mathbf{f}_a^{F:p} +  \mathbf{f}_a^{F:v},
\end{equation} 
where inter-particle averaged first Piola-Kirchhoff stress $\tilde{\mathbb{P}}$ 
is defined as
\begin{equation}
\tilde{\mathbb{P}}_{ab} = \frac{1}{2} \left( \mathbb{P}_a \mathbb{B}_a^0 + \mathbb{P}_b \mathbb{B}_b^0 \right). 
\end{equation}
Also $\mathbf{f}_a^{F:p}$ and $\mathbf{f}_a^{F:v}$ correspond to the fluid pressure and viscous forces acting on the solid particle $a$, respectively. 

In fluid-strucutre coupling, 
the surrounding solid structure is behaving as a moving boundary for fluid,
and the no-slip boundary condition is imposed at the fluid-structure interface.
Following Han and Hu \cite{han2018sph} and Adami et al. \cite{adami2012generalized}, 
the interaction forces $\mathbf{f}_i^{S:p}$ and $\mathbf{f}_i^{S:v}$ acting on a fluid particle $i$, 
due to the presence of the neighboring solid particle $a$, 
can be obtained as
\begin{equation}\label{S:p}
\mathbf{f}_i^{S:p} = - 2  \sum_a V_i V_a \frac{p_i \rho^d_a+ p^d_a \rho_i}{\rho_i + \rho^d_a} \nabla_i W(\mathbf{r}_{ia}, h ), 
\end{equation}
and 
\begin{equation}\label{S:v}
\mathbf{f}_i^{S:v} = 2\sum_a  \eta V_i V_a \frac{\mathbf{v}_i - \mathbf{v}^d_a}{|\mathbf{r}_{ia}| + 0.01h} \frac{\partial W(\mathbf{r}_{ia}, h )}{\partial {{r}_{ia}}} .
\end{equation}
Here, the imaginary pressure $p_a^d$ and velocity $\mathbf{v}_a^d$ are defined by
\begin{equation}\label{fs-coupling}
\begin{cases}
p_a^d = p_i + \rho_i max(0, (\mathbf{g} - {\frac{\text{d}\mathbf{v}_a}{\text{d}t}}) \cdot \mathbf{n}^S) (\mathbf{r}_{ia} \cdot \mathbf{n}^S) \\
\mathbf{v}_a^d = 2\mathbf{v}_i - \mathbf{v}_a
\end{cases},
\end{equation}
where $\mathbf{n}^S$ denotes the surface normal direction of the solid structure, 
and the imaginary particle density $\rho_a^d$ is calculated through 
the EoS presented in Eq. \eqref{eqeos} \cite{adami2012generalized}. 
Accordingly, the interaction forces $\mathbf{f}_a^{F:p}$ and $\mathbf{f}_a^{F:v}$ acting on a solid particle $a$ are given by
\begin{equation}\label{F:p}
\mathbf{f}_a^{F:p} = - 2  \sum_i V_a V_i \frac{p^d_a \rho_i +p_i \rho^d_a}{\rho_i + \rho^d_a} \nabla_a W(\mathbf{r}_{ai}, h ), 
\end{equation}
and 
\begin{equation}\label{F:v}
\mathbf{f}_a^{F:v} = 2\sum_i  \eta V_a V_i \frac{\mathbf{v}^d_a - \mathbf{v}_i}{|\mathbf{r}_{ai}| + 0.01h} \frac{\partial W(\mathbf{r}_{ai}, h )}{\partial {{r}_{ai}}} .
\end{equation}
The anti-symmetric property of the derivative of the kernel function will ensure the momentum conservation for each pair of interacting particles $i$ and $a$.  
For time integration, 
a single time step \cite{antoci2007numerical} 
\begin{equation}
 \Delta t = \min \left(0.25 \min(\frac{h}{c^F + |\mathbf{v}|_{max}}, \frac{{h}^2}{\eta}), 0.6 \min(\frac{h}{c^S + |\mathbf{v}|_{max}} ,
\sqrt{\frac{h}{|\frac{\text{d}\mathbf{v}}{\text{d}t}|_{max}}} ) \right) 
\end{equation}
is commonly used to ensure momentum conservation and force-calculation matching of fluid-structure interaction.  
Note that, 
this may lead to low computational efficiency 
as a very small time step can be chosen when the sound speed of the solid structure is significantly larger than that of the fluid. 
%
%
\section{Multi-resolution method}\label{sec:method}
In this section, 
we present a multi-resolution SPH method, 
where the fluid and solid equations are discretized by different spatio-temporal resolutions, 
to increase the computational efficiency. 
For this, different particle spacing, hence different smoothing lengths, 
and different time steps are utilized to discretize the fluid and solid equations. 
Note that, 
introducing different time steps may results in the issues of momentum conservation and force mismatch in fluid-structure coupling.  
To address these issues, 
we propose a position-based Verlet time integration scheme and 
introduce the time-averaged velocity and acceleration of solid particles for the calculation of interaction forces. 
\subsection{Multi-resolution discretization}
As different spatial resolutions are applied, 
the fluid discretizations of Eqs. \eqref{mass-conservation-f}  and \eqref{momentum-conservation-f} are modified as
\begin{equation} \label{fluid-model-mr-con}
\frac{\text{d} \rho_i}{\text{d} t}  = 2\rho_i \sum_{j}  V_j (\mathbf{v}_i - \widetilde{\mathbf{v}}_{ij}) \nabla_i W_{ij}^{h^F} ,
\end{equation}
and 
\begin{equation} \label{fluid-model-mr-mom}
m_i \frac{\text{d} \mathbf{v}_i}{\text{d} t}  =  - 2  \sum_j V_i V_j \widetilde{p}_{ij} \nabla_i W_{ij}^{h^F} + 
2\sum_j \eta V_i V_j \frac{{{\mathbf{v}}_{ij}}}{{{r}_{ij}}} \frac{\partial {{W}_{ij}^{h^F}}}{\partial {{r}_{ij}}} + \mathbf{g} +
\mathbf{f}_i^{S:p}\left(h^F\right) + \mathbf{f}_i^{S:v}\left(h^F\right),
\end{equation}
where $h^F$ represents the smoothing length used for fluid. 

Also, the discretization of solid momentum equation of Eq. \eqref{mom-sph} is modified to
\begin{equation}\label{solid-model-mr}
m_a \frac{\text{d} \mathbf{v}_a}{\text{d}t} = 2 \sum_b V_a V_b \tilde{\mathbb{P}}_{ab} \nabla_a^0 W_{ab}^{h^S} +\mathbf{g} + \mathbf{f}_a^{F:p}\left(h^F\right) + \mathbf{f}_a^{F:v}\left(h^F\right), 
\end{equation} 
where $h^S$ denotes the smoothing length used for solid structure. 
Note that,
for the calculation of fluid-structure interaction forces, 
we use $h^F$ with assumption $h^F \geqslant h^S$. 
This will ensure that a fluid particle $i$ can be searched and tagged as a neighboring particle 
of a solid particle $a$ which is located in the neighborhood of particle $i$. 
In detail, 
the forces $\mathbf{f}_i^{S:p}\left(h^F\right)$ and $\mathbf{f}_i^{S:v}\left(h^F\right)$ of Eqs. \eqref{S:p} and \eqref{F:v} are modified to
\begin{equation}\label{fs-force-mr-sp}
\mathbf{f}_i^{S:p}\left(h^F\right) =  - 2  \sum_a V_i V_a \frac{p_i \rho_a^d + p^d_a \rho_i}{\rho_i + \rho^d_a} \nabla_i W(\mathbf{r}_{ia}, h^F ),
\end{equation}
and
\begin{equation}\label{fs-force-mr-sv}
\mathbf{f}_i^{S:v}\left(h^F\right)= 2\sum_a  \eta V_i V_a \frac{\mathbf{v}_i - \mathbf{v}^d_a}{|\mathbf{r}_{ia}| + 0.01h} \frac{\partial W(\mathbf{r}_{ia}, h^F )}{\partial {{r}_{ia}}},
\end{equation}
Accordingly, 
the fluid forces exerting on the solid structure $\mathbf{f}_a^{F:p}\left(h^F\right) $ and $\mathbf{f}_a^{F:v}\left(h^F\right)$ can be obtained straightforwardly. 
\subsection{Multiple time steps}
\label{mr-time-step}
Here, the dual-criteria time-stepping method \cite{zhang2019dual} is employed for integrating the fluid equations. 
Two time-step criteria, 
namely the advection criterion $\Delta t_{ad}^F$ and the acoustic criterion $\Delta t_{ac}^F$, 
respectively, are 
\begin{equation}\label{dtf-advection}
\Delta t_{ad}^F   =  0.25 \min\left(\frac{h^F}{|\mathbf{v}|_{max}}, \frac{{h^F}^2}{\eta}\right),
\end{equation}
and 
\begin{equation}\label{dt-relax}
\Delta t_{ac}^F   = 0.6 \min \left( \frac{h^F}{c^F + |\mathbf{v}|_{max}} \right) .
\end{equation}
Here, 
the advection criterion controls the updating of the particle-neighbor list and the corresponding computation of the kernel values and gradients,
and the acoustic criterion determines the time integration of the particle motion. 
The time-step criterion for the solid structure is given as
\begin{equation}\label{dts-advection}
\Delta t^S   =  0.6 \min\left(\frac{h^S}{c^S + |\mathbf{v}|_{max}},
\sqrt{\frac{h^S}{|\frac{\text{d}\mathbf{v}}{\text{d}t}|_{max}}} \right) .
\end{equation}
Generally, 
$\Delta t^S < \Delta t_{ac}^F$, 
due to the fact that $c^S >  c^F$.
Other than choosing $\Delta t^S$ as the single time step for both fluid and structure, 
we carry out the structure time integration $\kappa = [\frac{\Delta t_{ac}^F}{\Delta t^S}] + 1$ times, 
where $[\cdot ]$ represents the integer operation, during one acoustic time step of fluid integration. 
As different time steps are applied in the integration of fluid and solid equations, 
the issue of force mismatch in the fluid-structure interaction may be encountered. 
That is, 
in the imaginary pressure and velocity calculation, 
the velocity and acceleration of solid particles in Eq. \eqref{fs-coupling} may present several different values updated after each $\Delta t^S$. 
Another issue is that the momentum conservation in the fluid and structure coupling may be violated. 

To address the force-calculation mismatch, 
we redefine the imaginary pressure $p_a^d$ and velocity $\mathbf{v}_a^d$ in Eqs. \eqref{fs-force-mr-sp} and \eqref{fs-force-mr-sv} as
\begin{equation} \label{fs-coupling-mr }
\begin{cases}
p_a^d = p_i + \rho_i max(0, (\mathbf{g} - \widetilde{\frac{\text{d} \mathbf{v}_a}{\text{d}t}}) \cdot \mathbf{n}^S) (\mathbf{r}_{ia} \cdot \mathbf{n}^S) \\
\mathbf{v}_a^d = 2 \mathbf{v}_i  - \widetilde{\mathbf{v}}_a
\end{cases}, 
\end{equation}
where \eqref{fs-coupling-mr } $\widetilde{\mathbf{v}}_a$ and $\widetilde{\frac{d\mathbf{v}_a}{dt}}$ 
represents the single averaged velocity and acceleration of solid particles during a fluid acoustic time step. 
\subsection{Position-based Verlet scheme}\label{sec:postion-verlet}
\begin{figure}[htb!]
	\centering
	\includegraphics[trim = 1cm 0cm 2cm 0cm, clip,width=0.95\textwidth]{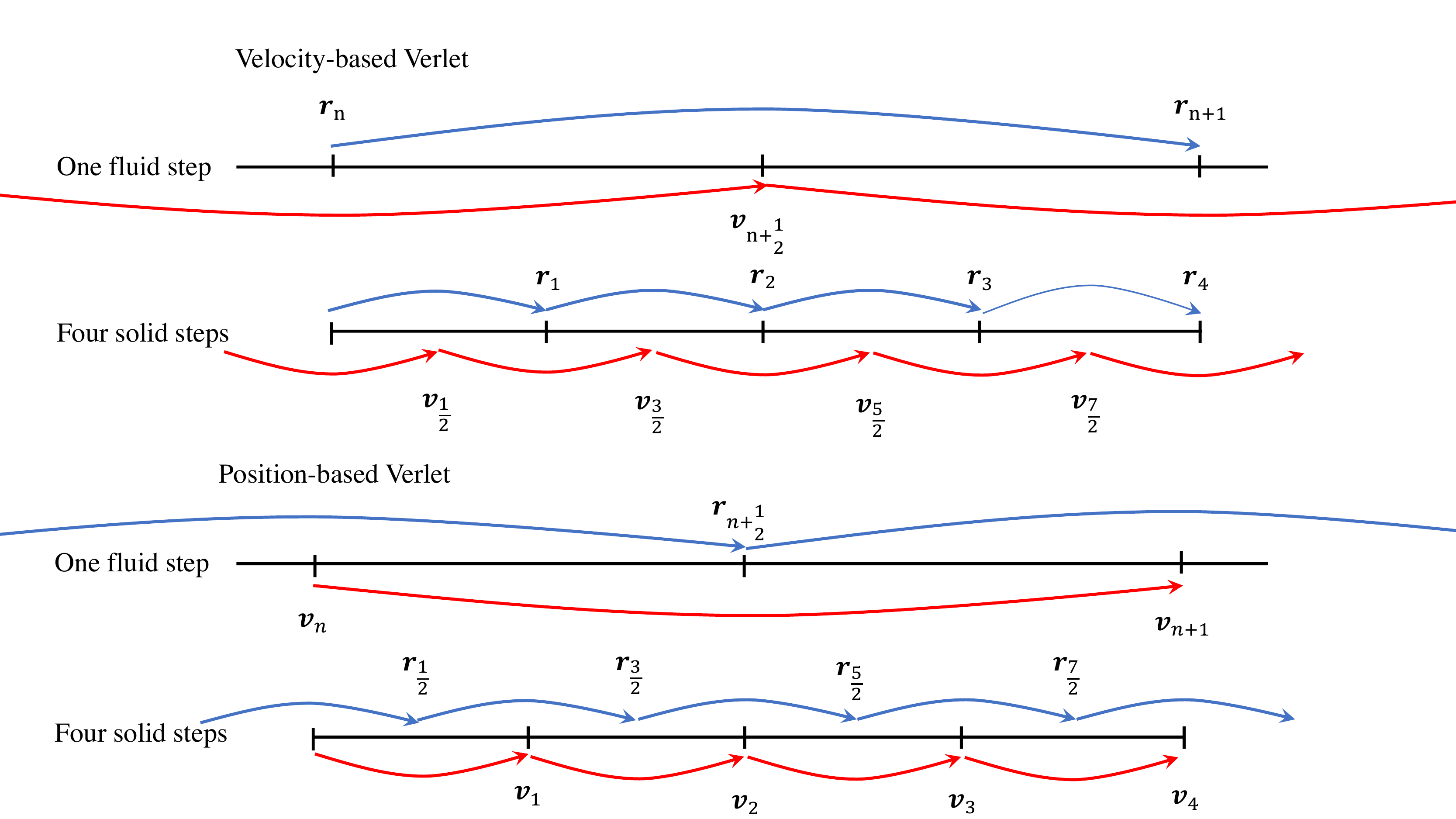}
	\caption{Sketch of velocity- and position-based Verlet scheme with assumption that $\kappa = 4$.}
	\label{figs:verletsetup}
\end{figure}

Instead of starting with a half step for velocity  
followed by a full step for position and another half step for velocity 
as in the velocity-based Verlet scheme \cite{adami2012generalized}, 
the position-based Verlet does the opposite: 
a half step for position followed by a full step for velocity and another half step for position. 
Figure \ref{figs:verletsetup} depicts the velocity- and position-based Verlet schemes assuming that $\kappa = 4$ for the integration of fluid and solid equations. 

Here, we denote the values at the beginning of a fluid acoustic time step by superscript $n$, 
at the mid-point by $n + \frac{1}{2}$ and eventually at the end of the time-step by $n + 1$.
At first, the integration of the fluid is conducted as
\begin{equation}\label{verlet-first-half}
\begin{cases}
\rho_i^{n + \frac{1}{2}} = \rho_i^n + \frac{1}{2}\Delta t_{ac}^F  \frac{d \rho_i}{dt}\\
\mathbf{r}_i^{n + \frac{1}{2}} = \mathbf{r}_i^n + \frac{1}{2} \Delta t_{ac}^F {\mathbf{v}_i}^{n}
\end{cases}, 
\end{equation}
by updating the density and position fields into the mid-point. 
Then particle velocity is updated to the new time step in the following form
\begin{equation}\label{verlet-first-mediate}
\mathbf{v}_i^{n + 1} = \mathbf{v}_i^{n} +  \Delta t_{ac}^F  \frac{d \mathbf{v}_i}{dt}. 
\end{equation}
Finally, the position and density of fluid particles are updated to the new time step by 
\begin{equation}\label{verlet-first-final}
\begin{cases}
\mathbf{r}_i^{n + 1} = \mathbf{r}_i^ {n + \frac{1}{2}} +  \frac{1}{2} \Delta t_{ac}^F {\mathbf{v}_i} \\
\rho_i^{n + 1} = \rho_i^{n + \frac{1}{2}} + \frac{1}{2} \Delta t_{ac}^F \frac{d \rho_i}{dt}
\end{cases}. 
\end{equation}
At this point, 
the pressure and viscous forces induced acceleration of solid particles are considered as constant during the integration for solid particles. 
For convenience of explanation, index $\varkappa = 0, 1, ...,  \kappa-1 $ is used to denote a integration step for solid particles.  
With the position-based Verlet scheme, 
the deformation tensor, density and particle position are updated to the midpoint as 
\begin{equation}\label{verlet-first-half-solid}
\begin{cases}
\mathbb{F}_a^{\varkappa + \frac{1}{2}} = \mathbb{F}_a^{\varkappa} + \frac{1}{2} \Delta t^S \frac{\text{d} \mathbb{F}_a}{\text{d}t}\\
\rho_a^{\varkappa + \frac{1}{2}} = \rho_a^0 \frac{1}{J} \\
\mathbf{r}_a^{\varkappa + \frac{1}{2}} = \mathbf{r}_a^{\varkappa} + \frac{1}{2} \Delta t^S {\mathbf{v}_a}
\end{cases}. 
\end{equation}
Then the velocity is updated by
\begin{equation}\label{verlet-first-mediate-solid}
\mathbf{v}_a^{\varkappa + 1} = \mathbf{v}_a^{\varkappa} +  \Delta t^S  \frac{d \mathbf{v}_a}{dt}. 
\end{equation}
At final stage, the deformation tensor and position of solid particles are updated to the new time step of solid structure with 
\begin{equation}\label{verlet-first-final-solid}
\begin{cases}
\mathbb{F}_a^{\varkappa + 1} = \mathbb{F}_a^{\varkappa + \frac{1}{2}} + \frac{1}{2} \Delta t^S \frac{\text{d} \mathbb{F}_a}{\text{d}t}\\
\rho_a^{\varkappa + 1} = \rho_a^0 \frac{1}{J} \\
\mathbf{r}_a^{\varkappa + 1} = \mathbf{r}_a^{\varkappa + \frac{1}{2}} + \frac{1}{2} \Delta t^S {\mathbf{v}_a}^{\varkappa + 1}
\end{cases}. 
\end{equation}
Before starting the next fluid step, 
the time integration of solid particles, 
which is depicted by Eqs. \ref{verlet-first-half-solid} to \ref{verlet-first-final-solid}, 
is iterated $\kappa$ times.
The detailed algorithms of the present multi-resolution SPH method are given in Alg. \ref{alg:a}

In the position-based Verlet scheme, 
as the the velocity is updated only once in the current fluid acoustic time step criterion, 
time marching of momentum equations for fluid and solid are exactly consistent as the velocity marching interval $ (\Delta t_{ac}^F)_n= \sum_{\varkappa = 0}^{\kappa - 1} (\Delta t^S)_{\varkappa}$.
Therefore, 
the present position-based Verlet algorithm gives a way to achieve strict momentum conservation in fluid-structure coupling when multiple time steps is employed. 
In the contrast, 
the velocity-based Verlet scheme does not guarantee momentum conservation as 
$ 0.5 \left[  (\Delta t_{ac}^F)_{n-1} +  (\Delta t_{ac}^F)_{n} \right] \neq  0.5 \left[ \sum_{\varkappa = 0}^{\kappa -1} \big((\Delta t^S)_{\varkappa - 1} + (\Delta t^S)_{\varkappa} \big) \right]$, 
as shown in Figure \ref{figs:verletsetup}.
\begin{algorithm}[htb!]
	Setup parameters and initialize the simulation\;
	Compute the initial number density of fluid particles\;
	Compute the normal direction of structure surface\;
	Obtain the corrected configuration for the solid particles\;
	\While{simulation termination condition is not satified}
	{
		Get the fluid advection time step $\Delta t_{ad}^F$\;
		Reinitialize the fluid density\;
		Update the normal direction of the solid structure\;
		\While{$\sum \Delta t_{ac}^F$ $\leq$ $\Delta t_{ad}^F$}
		{
			Get the fluid acoustic time step $\Delta t_{ac}^F$\;
			Integrate the fluid equations with the position-based Verlet scheme\;
			Compute the pressure and viscous forces exerting on structure due to fluid\;
			\While{$\sum \Delta t^S$ $\leq$ $\Delta t_{ac}^F$}
			{
				Get the time step $\Delta t^S$ of solid structure\;
				Integrate the solid equations with the position-based Verlet scheme\;
			}
			Compute the time-averged velocity and acceleration of solid structure\;
		}
		Update the particle-neighbor list and kernel values and gradient for fluid particles \;
		Update the particle configuration between the fluid and solid particles \;
	}
	Terminate the simulation.
	\caption{The multi-resolution SPH method for FSI problems.}
	\label{alg:a}
\end{algorithm}
%
%
%
\section{Numerical examples}\label{sec:validation}
In this section, 
we first study two benchmarks, 
where the structures experience large deformation due to either flow-induced vibration or time-dependent water pressure, 
to validate the accuracy and computational efficiency of the present method.
Having the validation studies presented, 
we then demonstrate the versatility of the method for applications in bio-mechanical system, 
e.g., venous valve and passive flapping of fish-like body.
In all the following examples,
the $5th$-order Wendland smoothing kernel function with the smoothing lengths $h^F = 1.3dp^F$ and $h^S = 1.3dp^S$, 
where $dp^F$ and $dp^S$ represent the initial particle spacing for fluid and solid structure, 
are employed. 
\subsection{Flow-induced vibration of an elastic beam behind a cylinder}
\label{subsec:test-beamvibration}
\begin{figure}[htb!]
	\centering
	\includegraphics[trim = 3cm 5cm 4cm 5cm, clip,width=0.95\textwidth]{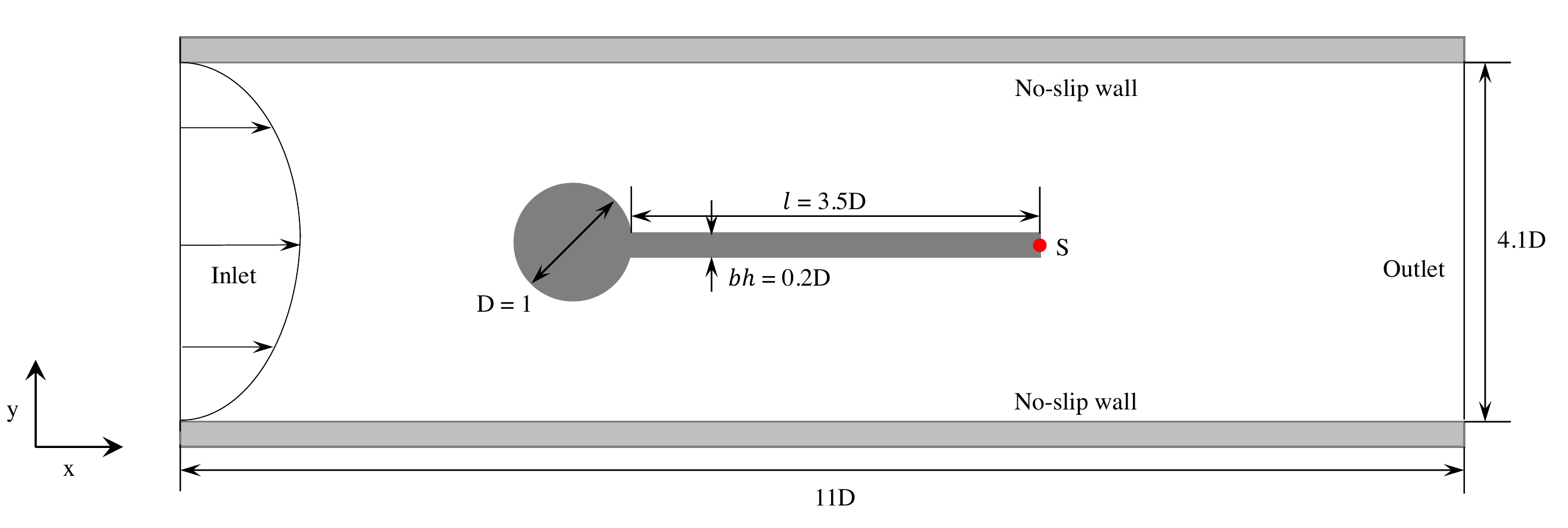}
	\caption{Sketch of the two dimensional flow-induced vibration of a flexible beam attached to a rigid cylinder with a trajectory sensor $S$ located at the free-end of the beam.
	}
	\label{figs:fsisetup}
\end{figure}

The first benchmark we consider herein 
is a two dimensional flow-induced vibration of a flexible beam attached to a rigid cylinder.
Following Turek and Hron \cite{turek2006proposal} and Han and Hu \cite{han2018sph},
the geometric parameters and basic setup of the problem are depicted in Figure \ref{figs:fsisetup}. 
The rigid cylinder is centered at $(2D, 2D)$ with $D = 1$ measured from the left bottom corner of the computational domain. 
No-slip boundary condition is imposed on the top and bottom walls, 
while inflow and outflow boundary conditions are employed at the left and 
right sides of the domain, respectively. 
The inflow condition is realized by imposing a velocity profile \cite{turek2006proposal,han2018sph} of 
$U(y) = 1.5 \overline{U}(t,y) (H-y)y / H^2$,
where $\overline{U} (t,y) =0.5U_0(1.0 - cos(0.5 \pi t))$ if $ t \leq t_s $ otherwise  $\overline{U}(t,y) = {U}_0$, 
with $U_0 = 1.0 $ and $t_s = 2.0$.
Similar to Refs. \cite{turek2006proposal, han2018sph}, 
the physical parameters are chosen to setup a challenging test, 
which consists of a quite flexible beam. 
The density ratio of the structure to the fluid is $\rho_S / \rho_F = 10$ and 
the Reynolds number $Re = \rho_{F}U_0 D/\eta = 100$.
The isotropic linear elastic model is employed with
the dimensionless Young's modulus $E^* = E/\rho_{F}U_0^2 = 1.4 \times 10^3$, 
and the Poisson ratio $\nu = 0.4$.
We consider four cases with different fluid-structure resolution ratios and  single or multiple time steps as shown in Table \ref{tab:fsi-cases} 
to study the computational efficiency and accuracy.
\begin{table}[htb!]
	\centering
	\caption{Flow-induced vibration of a flexible beam attached to a rigid cylinder: the setup of different cases. }
	\begin{tabular}{ ccccc}
		\hline
		Cases   & ${dp}_F/{dp}_S$ & ${dp}_S$     	& Time steps & Number of particles \\ 
		\hline
		I				& 1.0 & ${dp}_S = bh/4.0$     & Multiple  & 21664 \\
		\hline
		II 			& 1.0 & ${dp}_S = bh/8.0$     & Multiple  & 78200 \\
		\hline
		III	 		& 2.0 & ${dp}_S = bh/4.0$     & Multiple  & 6880 \\
		\hline
		IV 		  & 2.0 & ${dp}_S = bh/4.0$     & Single  & 6880 \\
		\hline	
	\end{tabular}
	\label{tab:fsi-cases}
\end{table}

\begin{figure}[htb!]
	\centering
	\includegraphics[trim = 6cm 0mm 15cm 0mm, clip,width=0.9\textwidth]{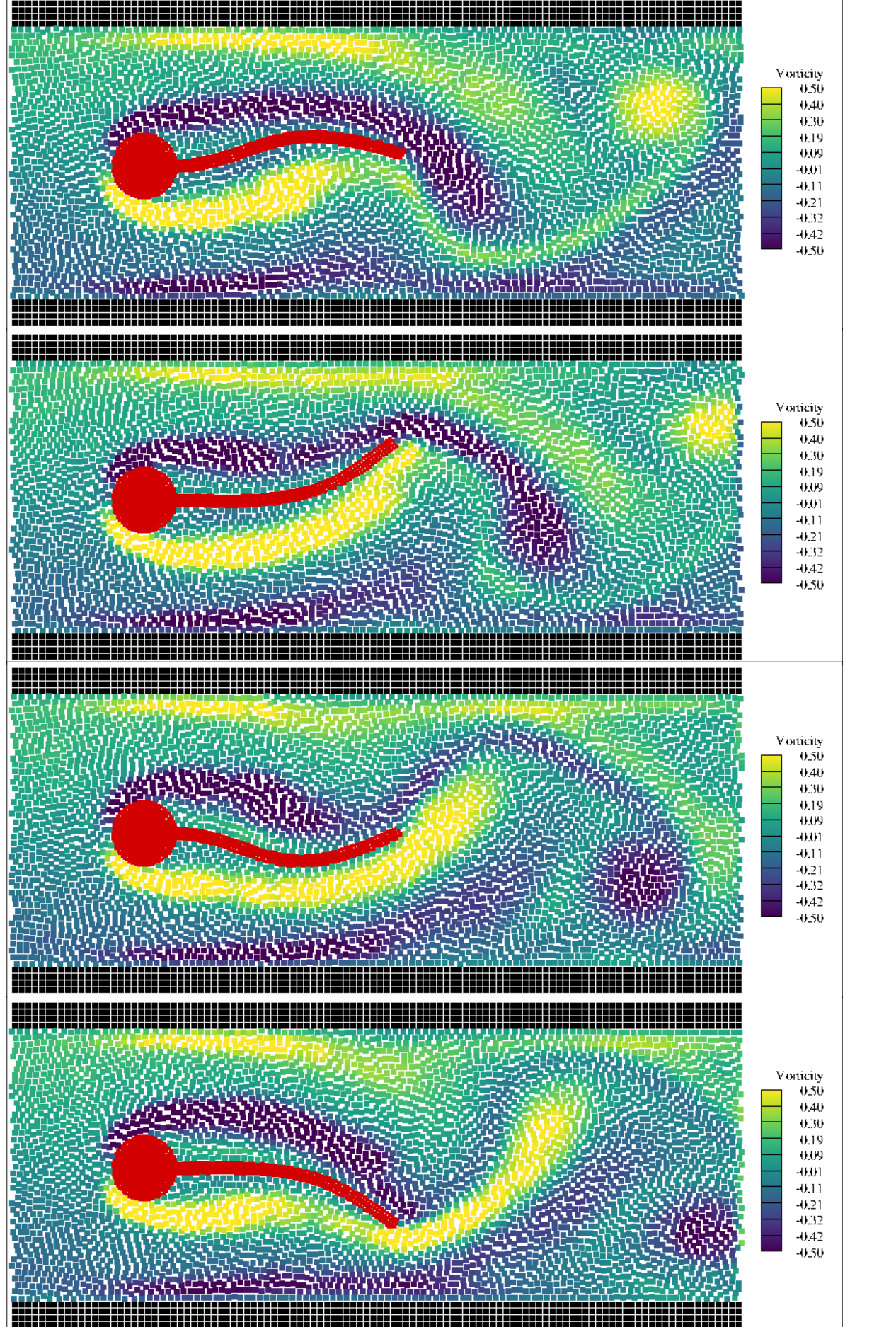}
	\caption{Flow-induced vibration of a beam attached to a cylinder with $d_F/d_S = 2.0$ (Case-III): the flow vorticity field and beam deformation at four time instants labeled by red dots in Figure \ref{figs:fsi-xy}.}
	\label{figs:fsi-particle} 
\end{figure}

\begin{figure}[htb!]
	\centering
	\includegraphics[trim = 1mm 1mm 1mm 1mm, clip,width=.85\textwidth]{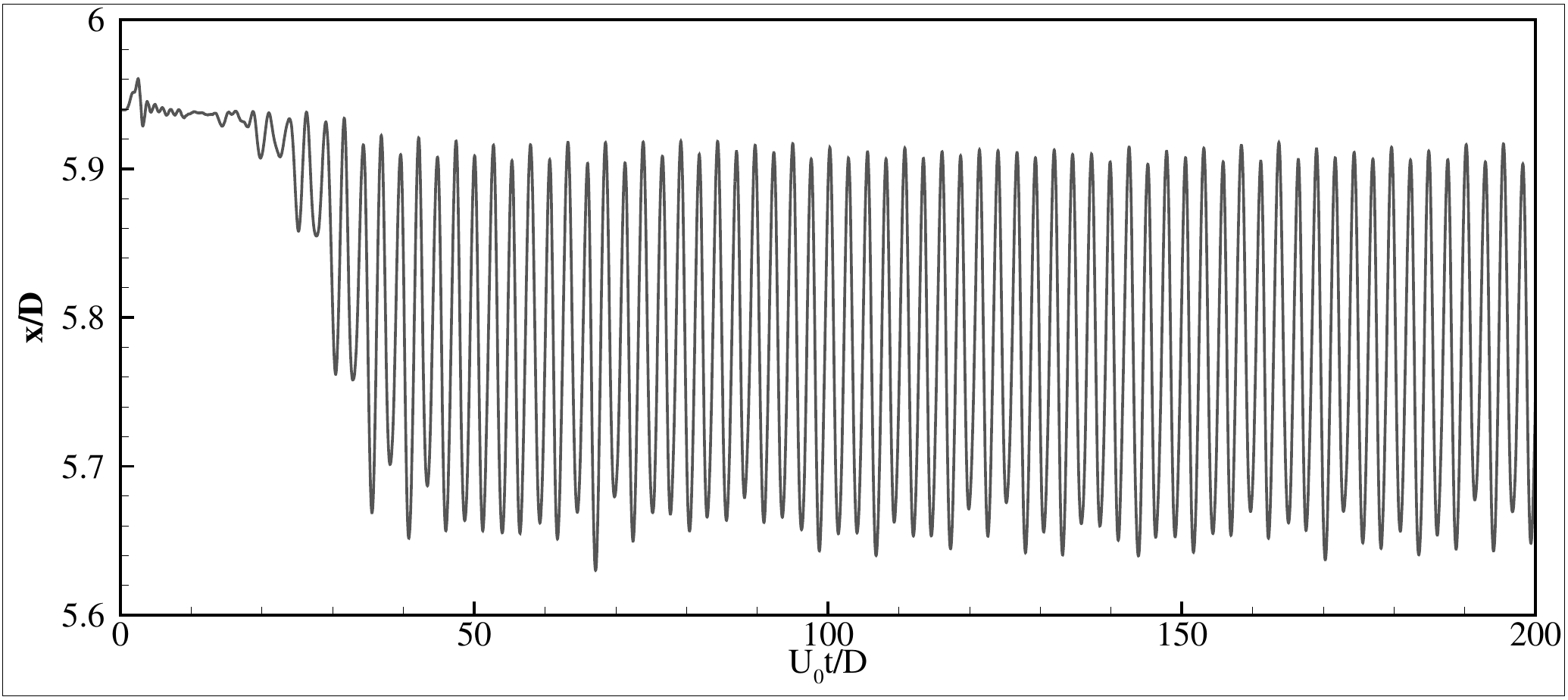}
	\includegraphics[trim = 1mm 1mm 1mm 1mm, clip,width=.85\textwidth]{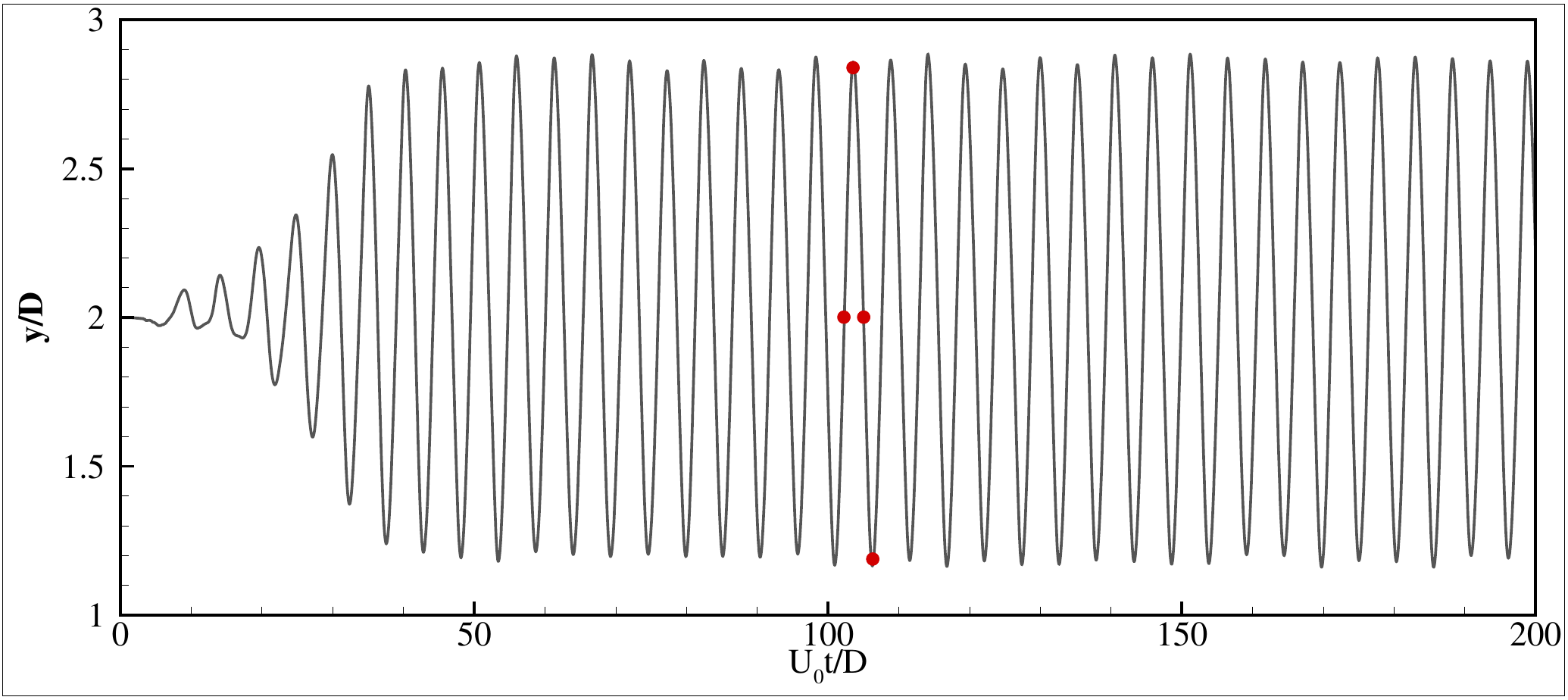}
	\includegraphics[trim = 1mm 1mm 1mm 1mm, clip,width=.85\textwidth]{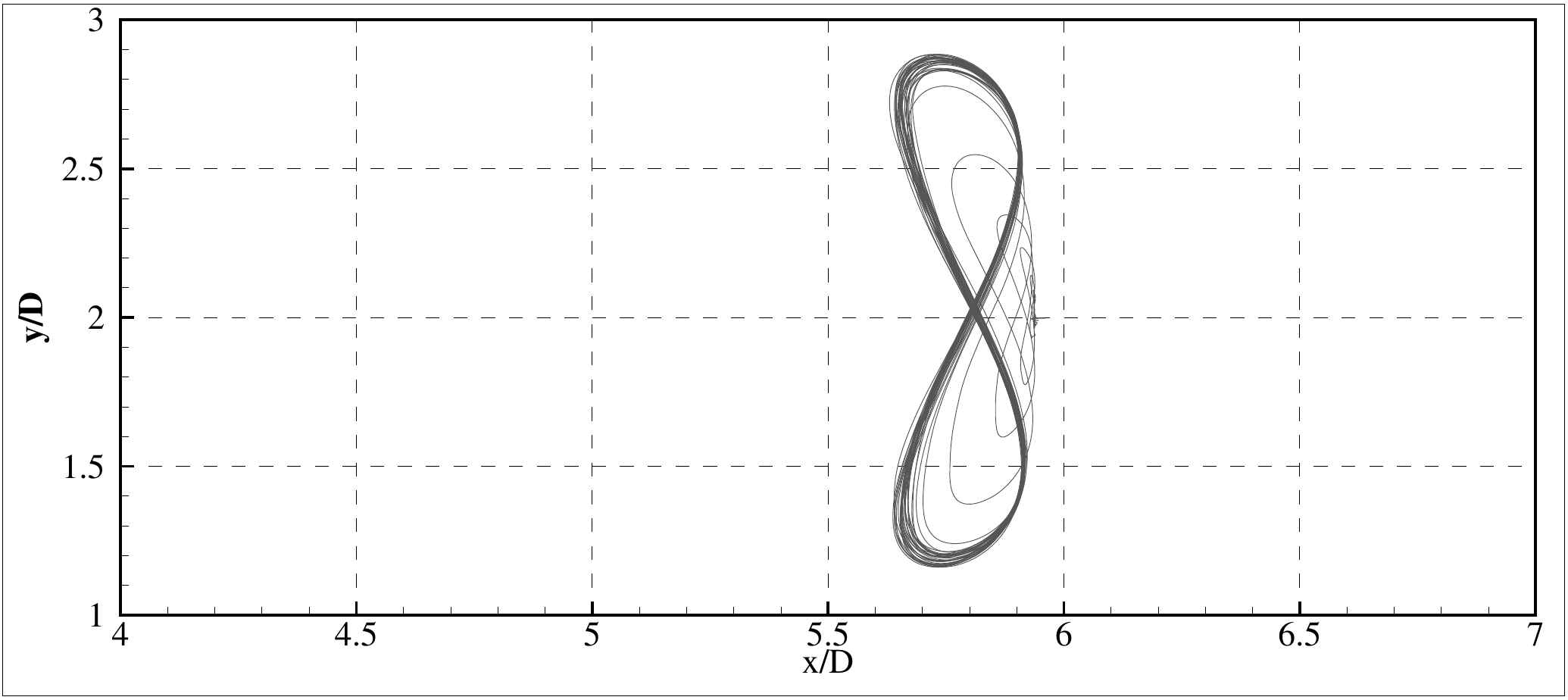}
	\caption{Flow-induced vibration of a beam attached to a cylinder (Case-III): amplitude of the displacement in $x$-direction (first panel), amplitude of the displacement in $y$-direction (second panel) and the trajectory (third panel) of point $S$.}
	\label{figs:fsi-xy}
\end{figure}

Figure \ref{figs:fsi-particle} shows the flow vorticity field and beam deformation for Case-III at different time instants
in a typical periodic movement when self-sustained oscillation is reached.
Good agreements with previous computational results \cite{turek2006proposal, han2018sph, bhardwaj2012benchmarking} are noted. 
Figure \ref{figs:fsi-xy} gives the time dependent position of the trajectory sensor $S$.
It is noted that the beam reaches a periodic self-sustained oscillation as time goes beyond a dimensionless time of 50.
The trajectory of point $S$ shows a typical Lissajous curve with a frequency ratio of $2:1$ between horizontal and vertical motions \cite{bhardwaj2012benchmarking}, 
implying a good agreement with the computational results of both Refs. \cite{bhardwaj2012benchmarking} and \cite{tian2014fluid}.
Table \ref{tab:fsi-data} gives the dimensionless amplitude of the oscillation in $y$ direction, 
the frequency obtained from the cases given in Table \ref{tab:fsi-cases}.
Compared with the previous results in the literature \cite{turek2006proposal,bhardwaj2012benchmarking,tian2014fluid,zhang2019dual}, 
where the frequency is predicted as $0.19$ and the amplitude is varied from $0.78$ to $0.92$, 
good agreement is noted. 
It can also be observed that with resolution ratio ${dp}_F/{dp}_S = 2.0$ and multiple time steps,  
one is able to obtain a speed up of $5.2$ without notable  compensation on numerical accuracy. 

To further evaluate the convergence of the present method,  
we performed  simulations with increasing the spatial resolution.
It is found that the amplitude and the frequency converge to $0.855$ and $0.189$, respectively, as noted in Table \ref{tab:fsi-data-convergence}. 
Compared with the converged results, 
the present method demonstrates good accuracy except with the slightly overestimated amplitude in $y$-direction. 
\begin{table}[htb!]
	\centering
	\caption{Flow-induced vibration of a beam attached to a cylinder: computational efficiency and predicted amplitude and frequency. 
		The computations are carried out on an Intel(R) Xeon(R) L5520 2.27GHz Desktop 
		computer with 24GiB RAM and Scientific Linux system (6.9). 
		To analyze the computational performance,  
		we evaluate the CPU wall-clock time for share memory parallelized computations (based on TBB library) until the dimensionless time of $200$. }
	\begin{tabular}{cccc}
		\hline
		Cases    & Amplitude in $y$-axis      & Frequency  & CPU time (s)\\ 
		\hline
		I & $0.89$       & $0.180$  &1297.7   \\ 
		II & $0.86$       & $0.184$  &10926.7   \\ 
		III & $0.86$       & $0.188$  &249.5    \\ 
		IV & $0.86$       & $0.188$  &451.4   \\ 
		\hline
	\end{tabular}
	\label{tab:fsi-data}
\end{table}
\begin{table}[htb!]
	\centering
	\caption{Flow-induced vibration of a beam attached to a cylinder: convergence study of Case-III.}
	\begin{tabular}{ ccc}
		\hline
		Spatial resolution   & Amplitude in $y$-axis      & Frequency  \\ 
		\hline
		$bh / {dp}_S = 4$ & $0.860$       & $0.188$     \\ 
		$bh / {dp}_S = 8$ & $0.856$       & $0.189$     \\ 
		$bh / {dp}_S = 16$ & $0.855$       & $0.189$     \\ 
		\hline
	\end{tabular}
	\label{tab:fsi-data-convergence}
\end{table}
%
\subsection{Dam-break flow through an elastic gate}\label{subsec:test-damplate}
\begin{figure}[htb!]
	\centering
	\includegraphics[trim = 0cm 0cm 0cm 0cm, clip,width=0.8\textwidth]{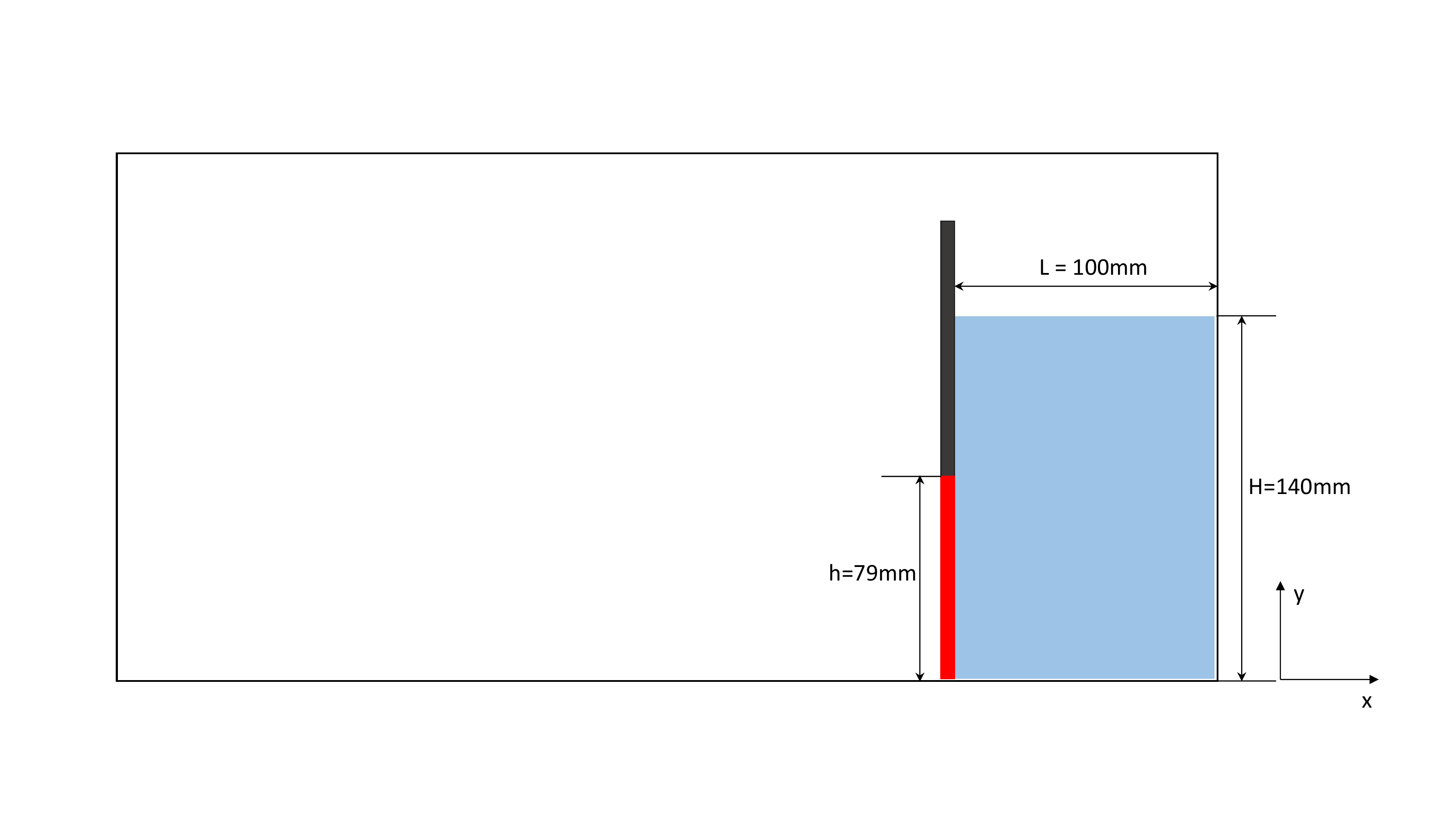}
	\caption{Sketch of the dam-break flow through an elastic gate which  has a width of $b = 5mm$. }
	\label{figs:damplate-setup}
\end{figure}

In this section, we consider the deformation of an elastic plate subjected to time-dependent water pressure, 
where free-surface flow is involved.  
Following the experiment study conducted by Antoci et al. \cite{antoci2007numerical}, 
the configuration is shown in Figure \ref{figs:damplate-setup} where an elastic plate is clamped at its upper end and is free at the lower one, 
interacting with a bulk of water initially confined in an open-air tank behind it.
Following Ref. \cite{antoci2007numerical}, 
the fluid flow is considered to be inviscid with the density $\rho_F = 1000 kg/m^3$.
The elastic plate corresponds to rubber and the linear isotropic model is employed with the properties of density $\rho_S = 1100kg/m^3$,
Young's modulus $E = 7.8MPa$ and Possion ratio $\nu = 0.47$.
Note that a more realistic Possion ratio $\nu$ is chosen herein compared with the numerical setup in Ref. \cite{antoci2007numerical} (therein $\nu = 0.4$).
The clamp condition is imposed to the upper end of the plate through constraints of the rigid base.
Here, 
similar to the previous problem, four cases as shown in Table \ref{tab:dam-cases} 
are computed to investigate the accuracy and computational efficiency. 
\begin{table}[htb!]
	\centering
	\caption{Dam-break flow through an elastic gate: basic setup of different cases.  }
	\begin{tabular}{ ccccc}
		\hline
		Cases   & ${dp}_F / {dp}_S$ & ${dp}_S$     	& Time steps & Number of particles \\ 
		\hline
		I				& 1.0 & ${dp}_S = b/4.0$     & Multiple	& 14144 \\
		\hline
		II 			& 1.0 & ${dp}_S = b/8.0$     & Multiple	& 47394 \\
		\hline
		III	 		& 2.0 & ${dp}_S = b/4.0$     & Multiple	& 5220 \\
		\hline
		IV 		  & 2.0 & ${dp}_S = b/4.0$     & Single  & 5220 \\
		\hline	
	\end{tabular}
	\label{tab:dam-cases}
\end{table}

\begin{figure}[htb!]
	\centering
	\includegraphics[trim = 1mm 1mm 1mm 10mm, clip,width=.495\textwidth]{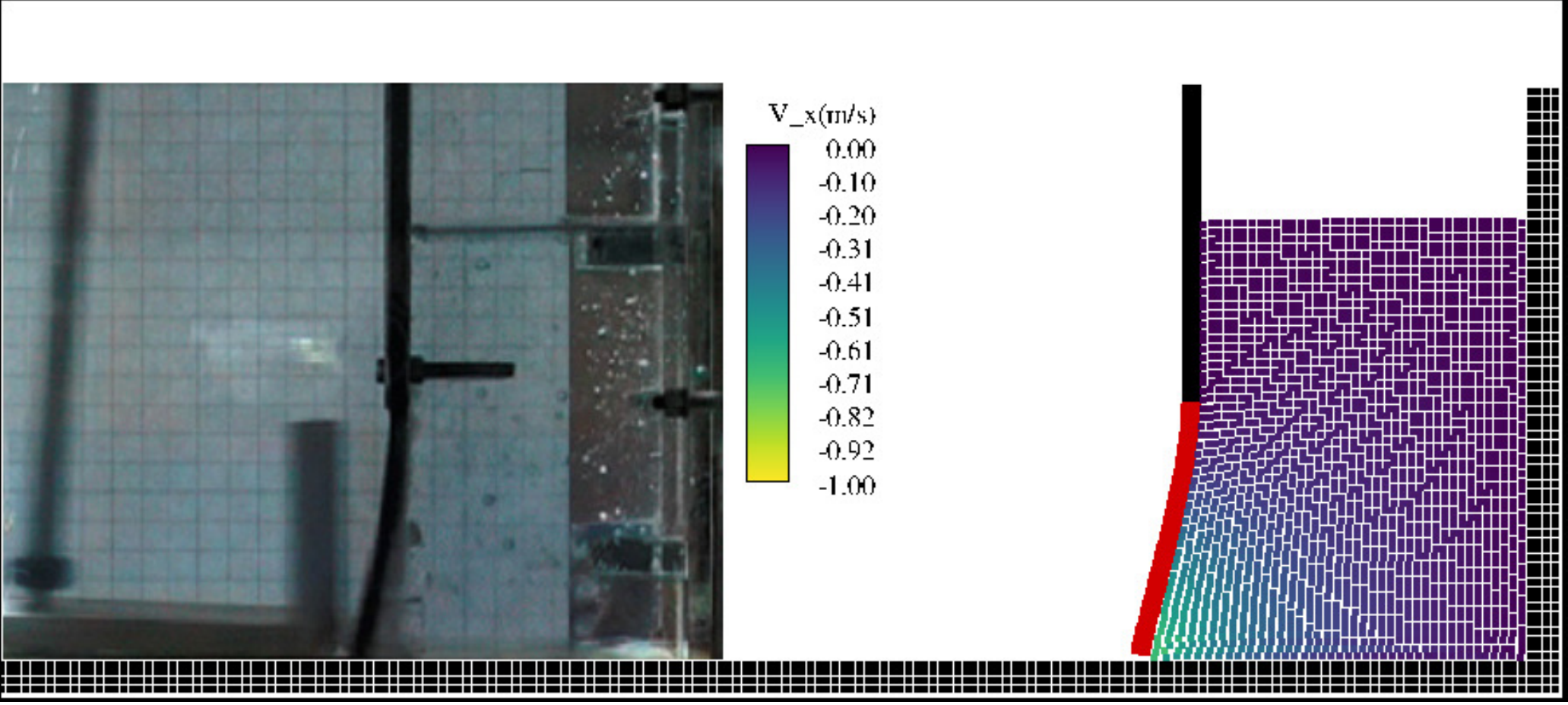}
	\includegraphics[trim = 1mm 1mm 1mm 10mm, clip,width=.495\textwidth]{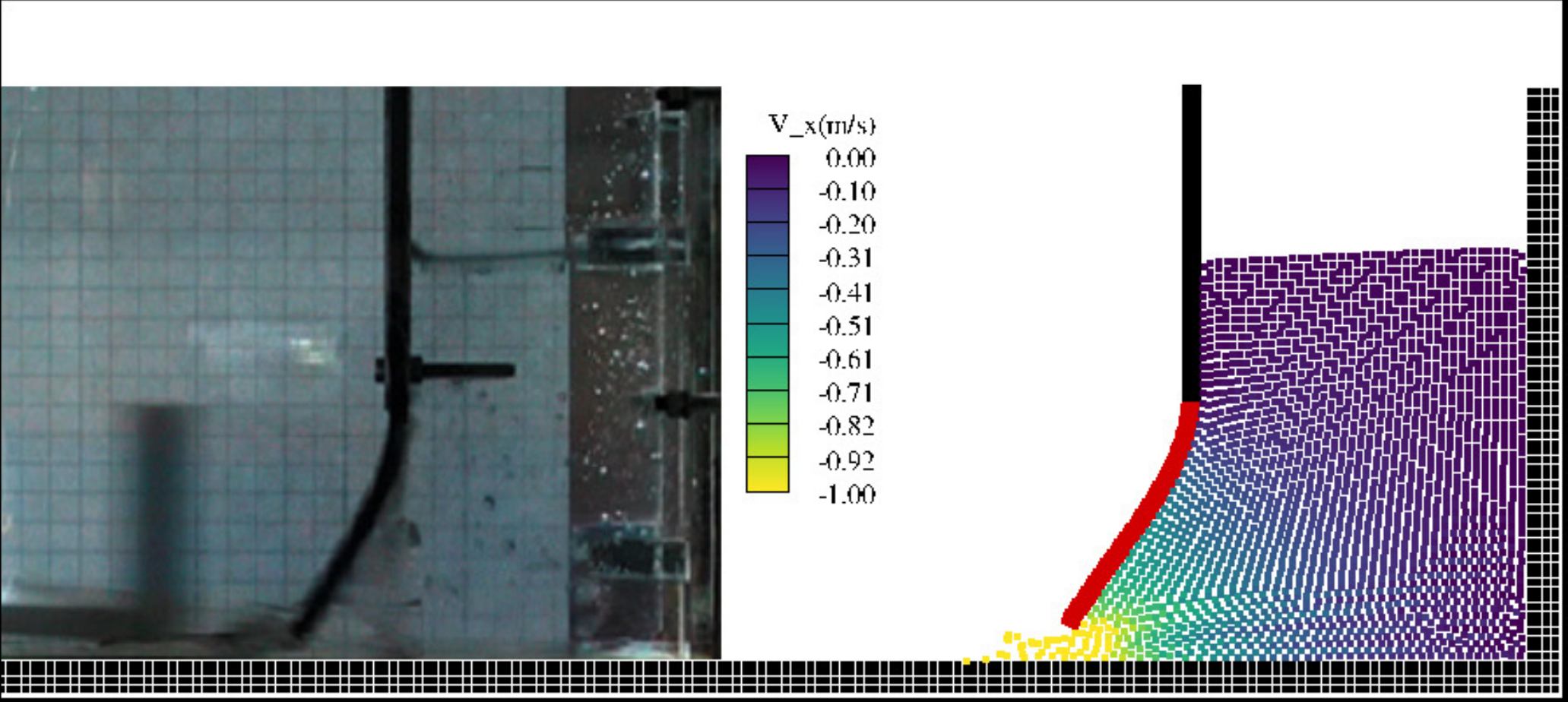}\\
	\includegraphics[trim = 1mm 1mm 1mm 10mm, clip,width=.495\textwidth]{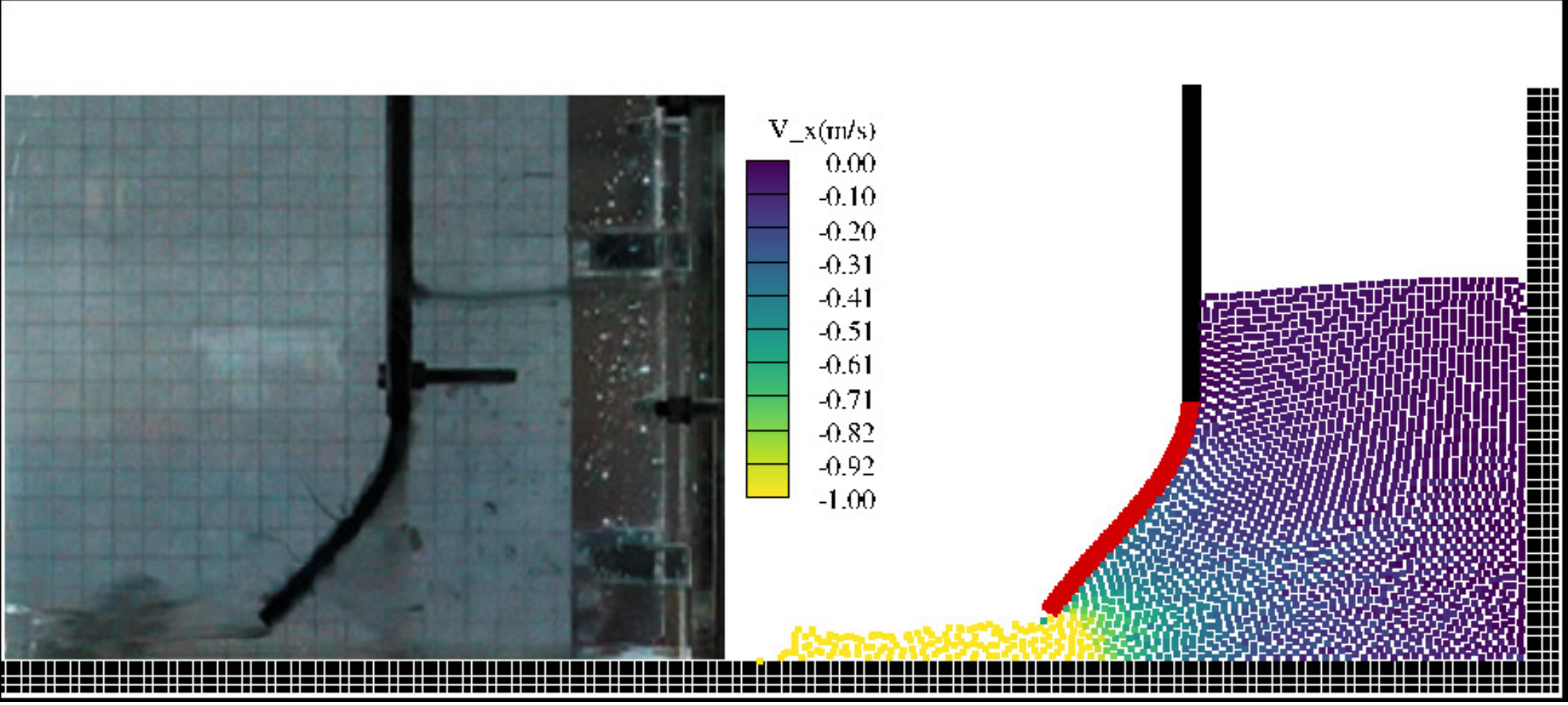}
	\includegraphics[trim = 1mm 1mm 1mm 10mm, clip,width=.495\textwidth]{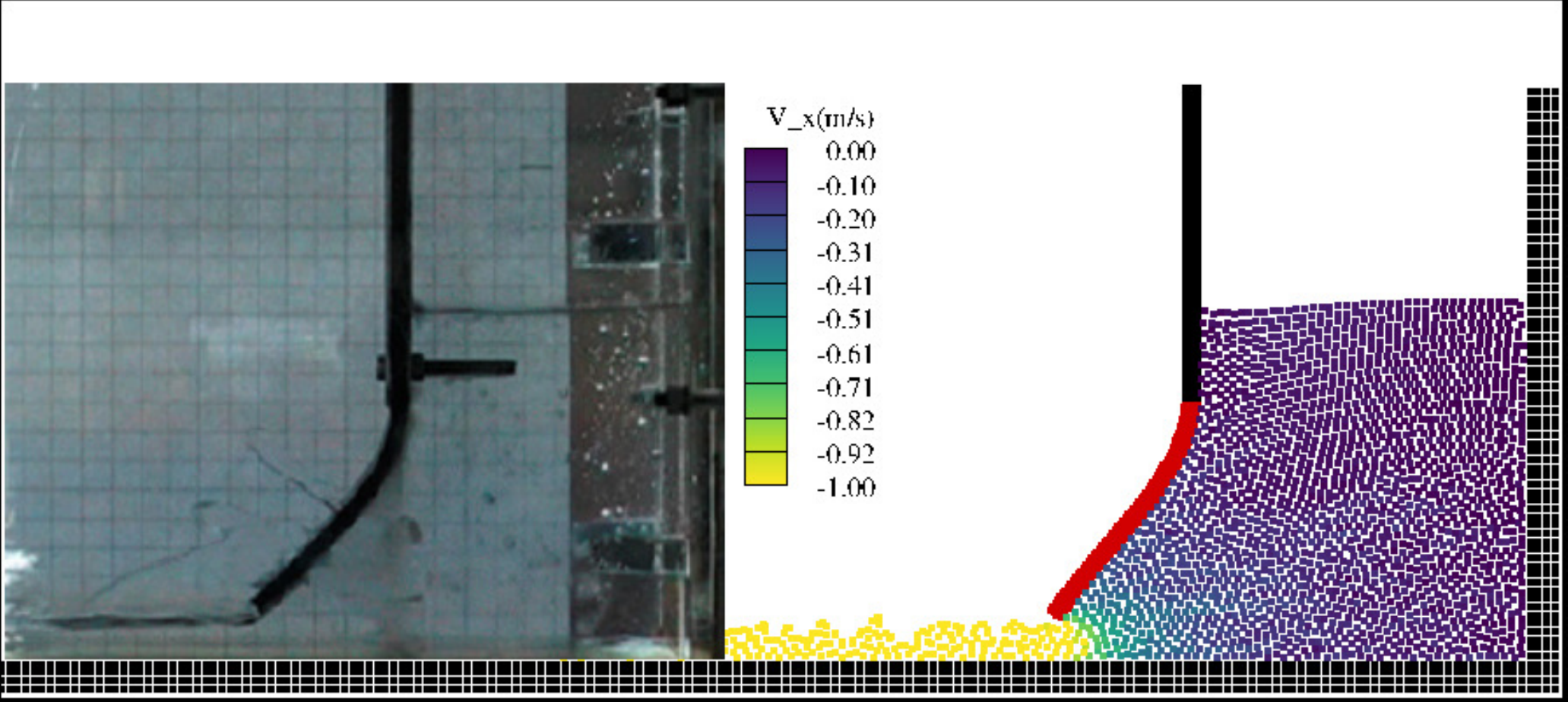}\\
	\includegraphics[trim = 1mm 1mm 1mm 10mm, clip,width=.495\textwidth]{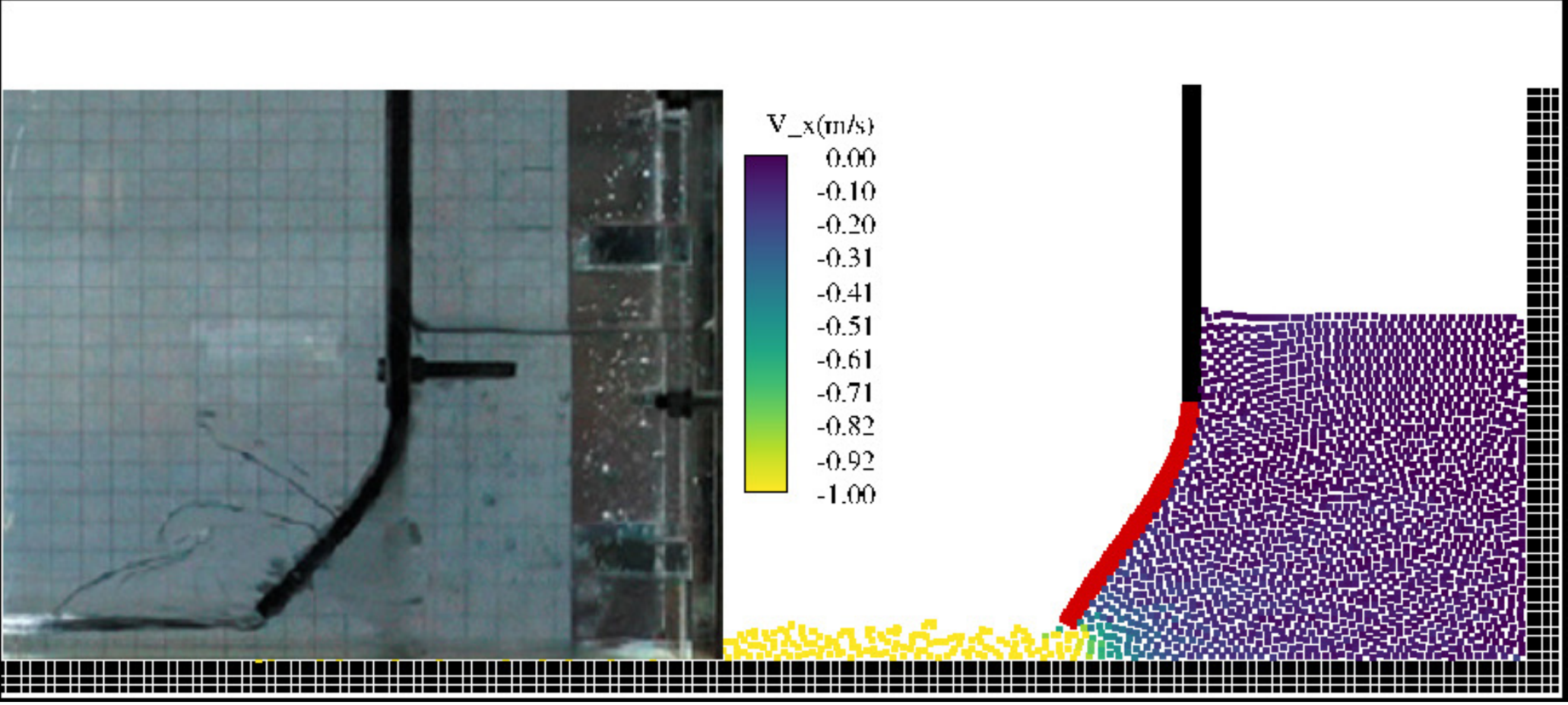}
	\includegraphics[trim = 1mm 1mm 1mm 10mm, clip,width=.495\textwidth]{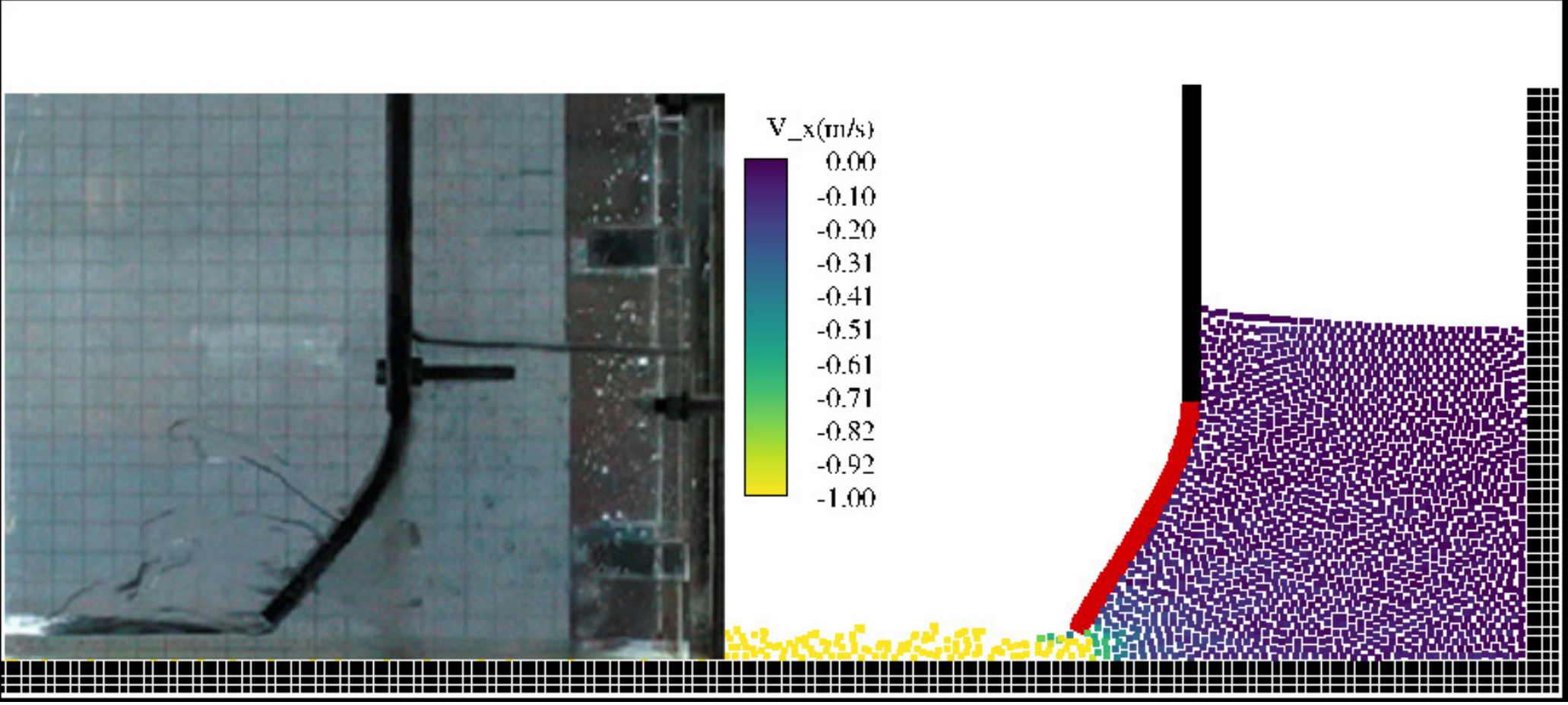}\\
	\includegraphics[trim = 1mm 1mm 1mm 10mm, clip,width=.495\textwidth]{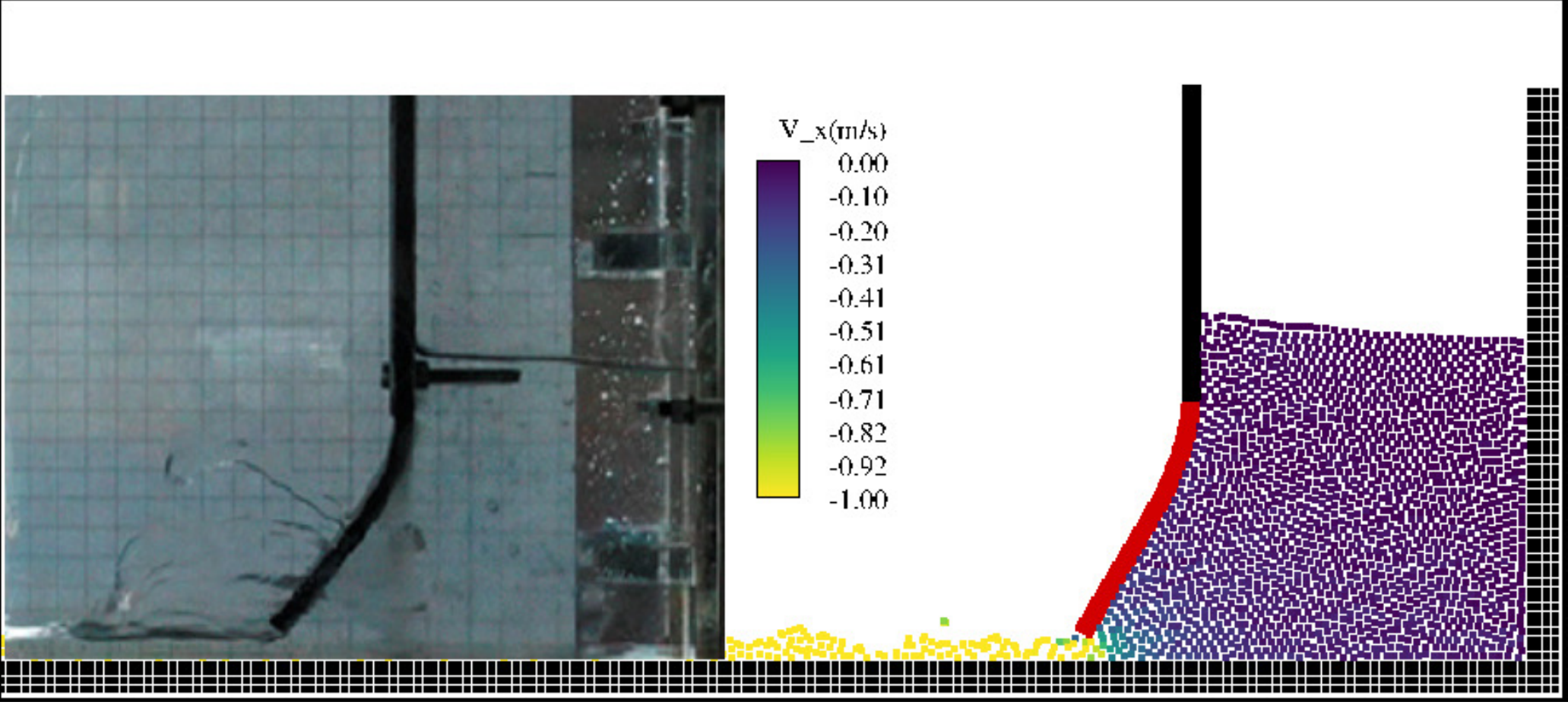}
	\includegraphics[trim = 1mm 1mm 1mm 10mm, clip,width=.495\textwidth]{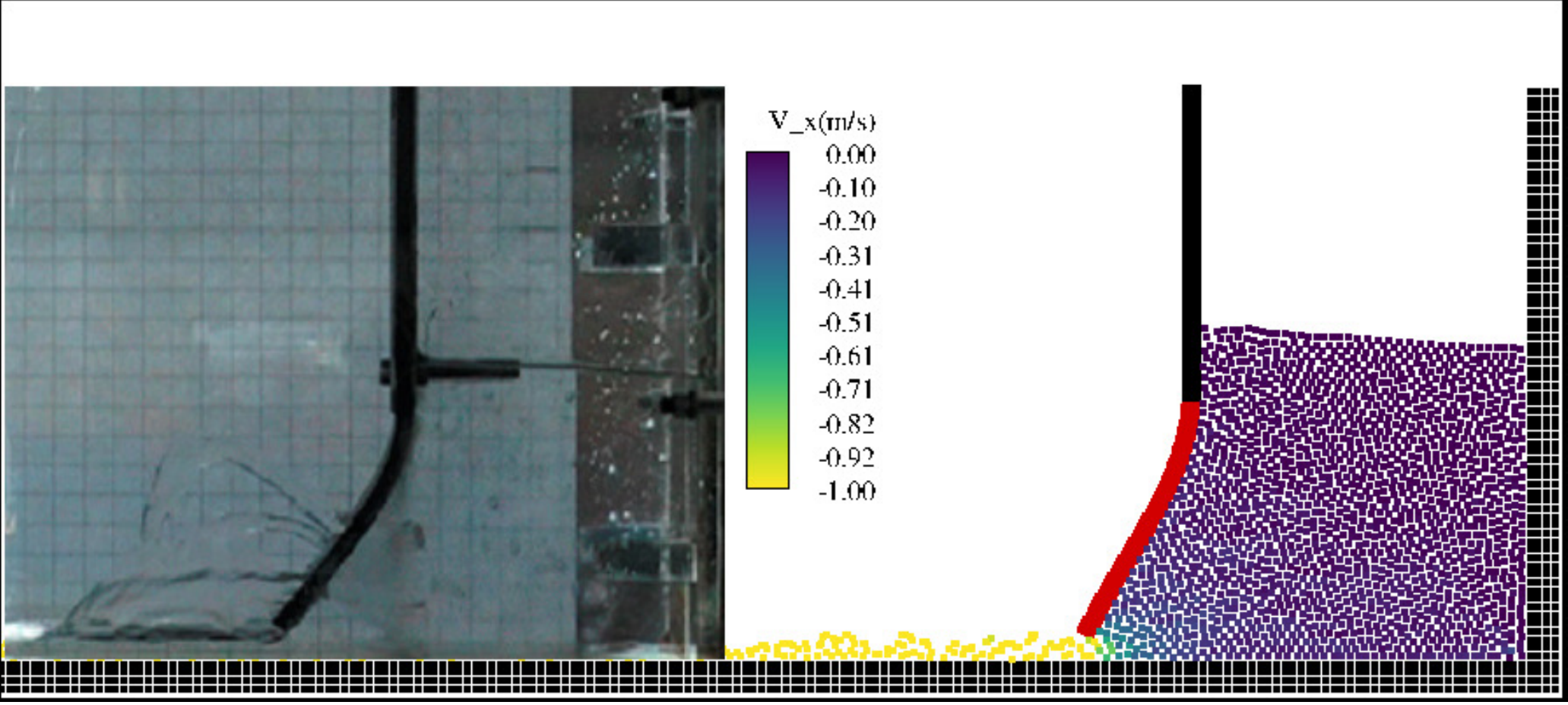}\\
	\caption{Dam-break flow through an elastic gate: 
		snapshots from Case-III 
		are compared against experimental frames presented by Antoci et al. \cite{antoci2007numerical}
		at every $0.04s$ starting from $t = 0.04s$.}
	\label{figs:dam-plate-surface}
\end{figure}

The comparison between the snapshots obtained from Case-III and the experiment presented by Antoci et al. \cite{antoci2007numerical} 
is illustrated in Figure \ref{figs:dam-plate-surface}.
These snapshots correspond to every $0.04s$, starting at time $t=0.04s $. 
It can be observed that the simulation shows a free-surface motion quite close to that of the experiment.
Note that, 
in experiment, there is a considerable leakage between the tank wall and the flexible plate resulting in splashes seen from time $t=0.08s$.
As the simulation herein is two dimensional, such splashes are not presented. 

\begin{figure}[htb!]
	\centering
	\includegraphics[trim = 1mm 1mm 1mm 1mm, clip,width=.85\textwidth]{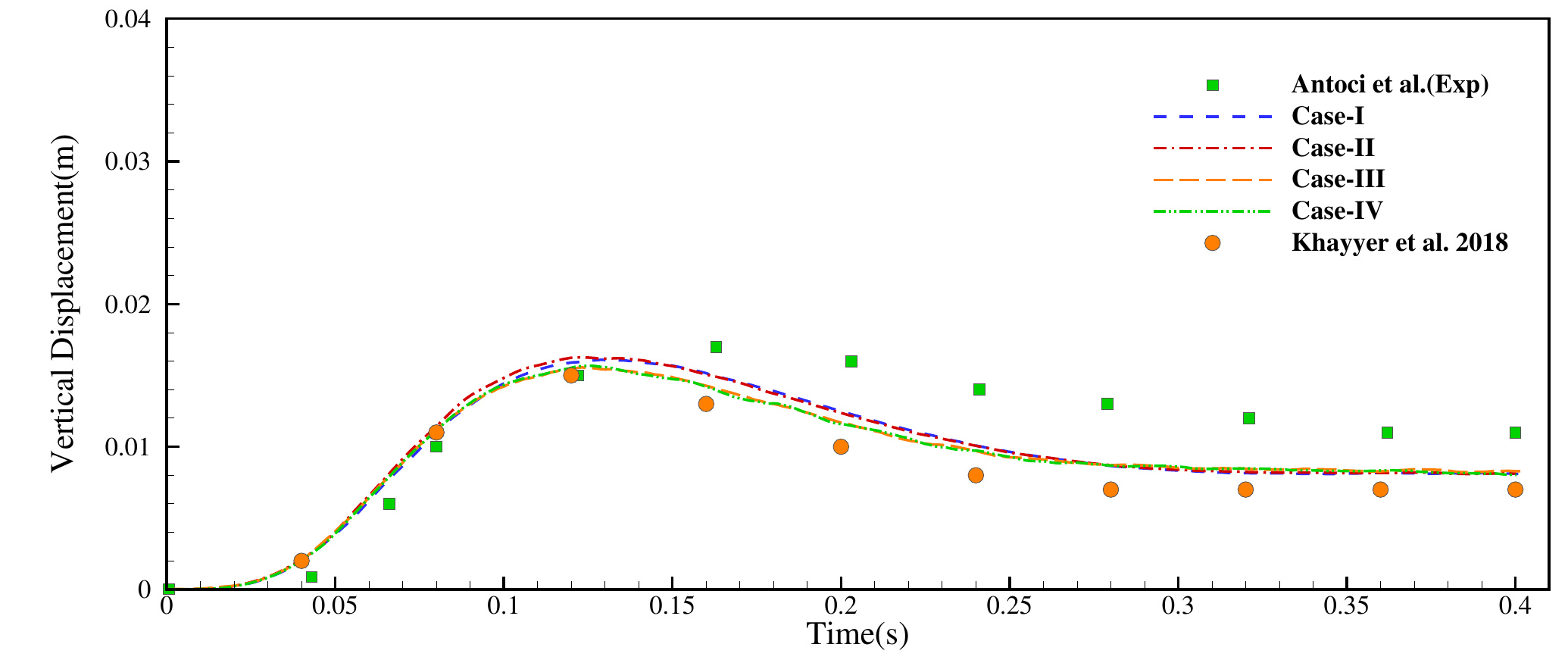}
	\includegraphics[trim = 1mm 1mm 1mm 1mm, clip,width=.85\textwidth]{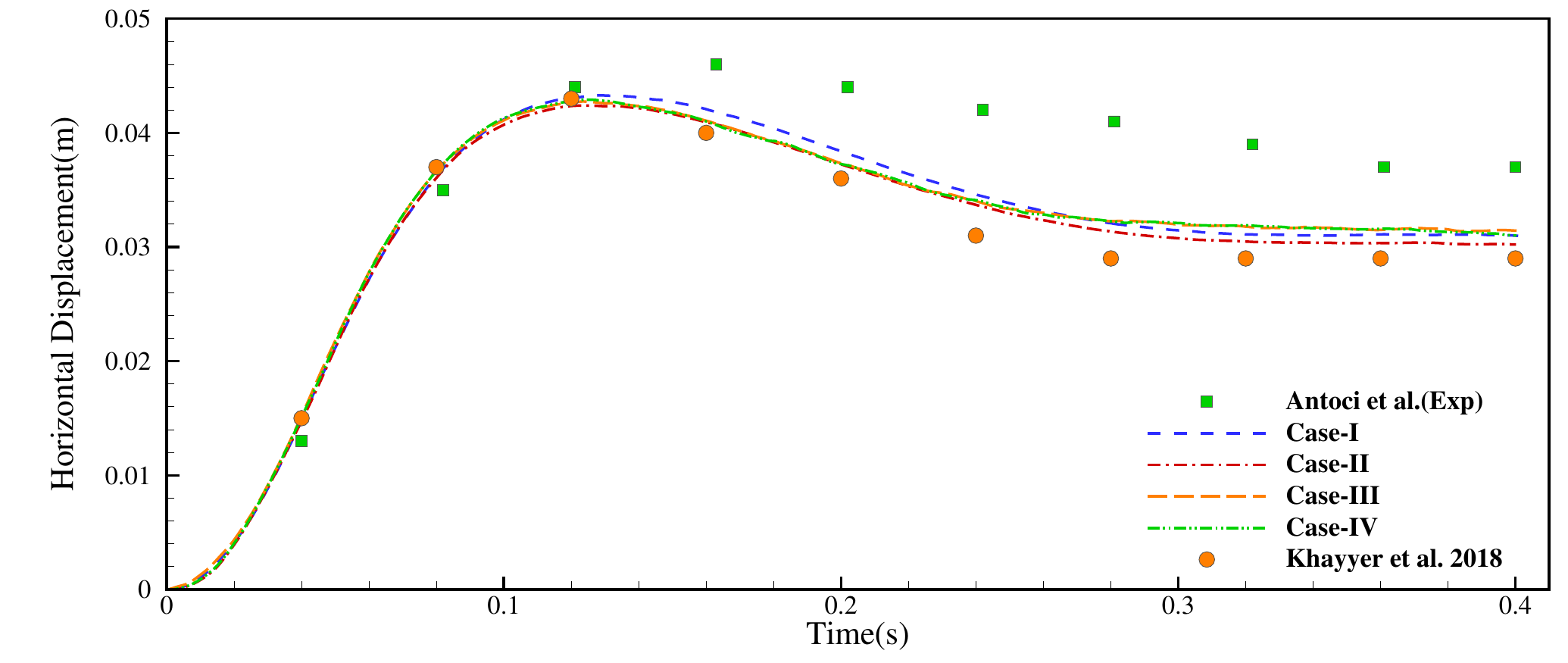}
	\caption{Dam-break flow through an elastic gate: 
		horizontal (top panel) 
		and vertical (bottom panel) displacement of the free end of the plate.
		Computational results are compared with experimental data provided by Antoci et al. \cite{antoci2007numerical} and previous numerical results of Ref. \cite{khayyer2018enhanced}. }
	\label{figs:dam-plate-data}
\end{figure}

The horizontal and vertical displacements of the free end of the plate 
obtained here are compared against the experiment \cite{antoci2007numerical} and previous numerical results in Khayyer et al. \cite{khayyer2018enhanced} in Figure \ref{figs:dam-plate-data}. 
For all the four cases, 
good agreements with experimental data and previous results are noted. 
In the early stages of the dambreak,
the horizontal displacement increases quickly due to the initial high water level and static pressure.
At time $t = 0.15s$, the plate reaches its maximum deformation with a horizontal displacement of approximately $0.43$.
As the water level and static pressure decreases, 
the plate gradually returns to its equilibrium state which is charaterized by the balance between the elastic force and the force introduced by the dynamic water pressure. 
It also worth noting that similar to previous simulations \cite{khayyer2018enhanced, zhang2019smoothed, yang2012free},  
small differences are observed between the present results and the experiment data during the closing period. 
These discrepancies can be associated to the material model employed in 
the simulations \cite{yang2012free}. 
Compared with the results with single resolution, 
the multi-resolution computation demonstrates similar fidelity, 
otherwise, shows optimized computational efficiency as reported in Table \ref{tab:dam-gate-cpu}. 
Another worth noting feature is that using multiple time steps shows very high computational performance when a large Possion ratio, 
which leads to a quite small $\Delta t^S$, is applied. 

\begin{figure}[htb!]
	\centering
	\includegraphics[trim = 1mm 2mm 2mm 1mm, clip,width=.85\textwidth]{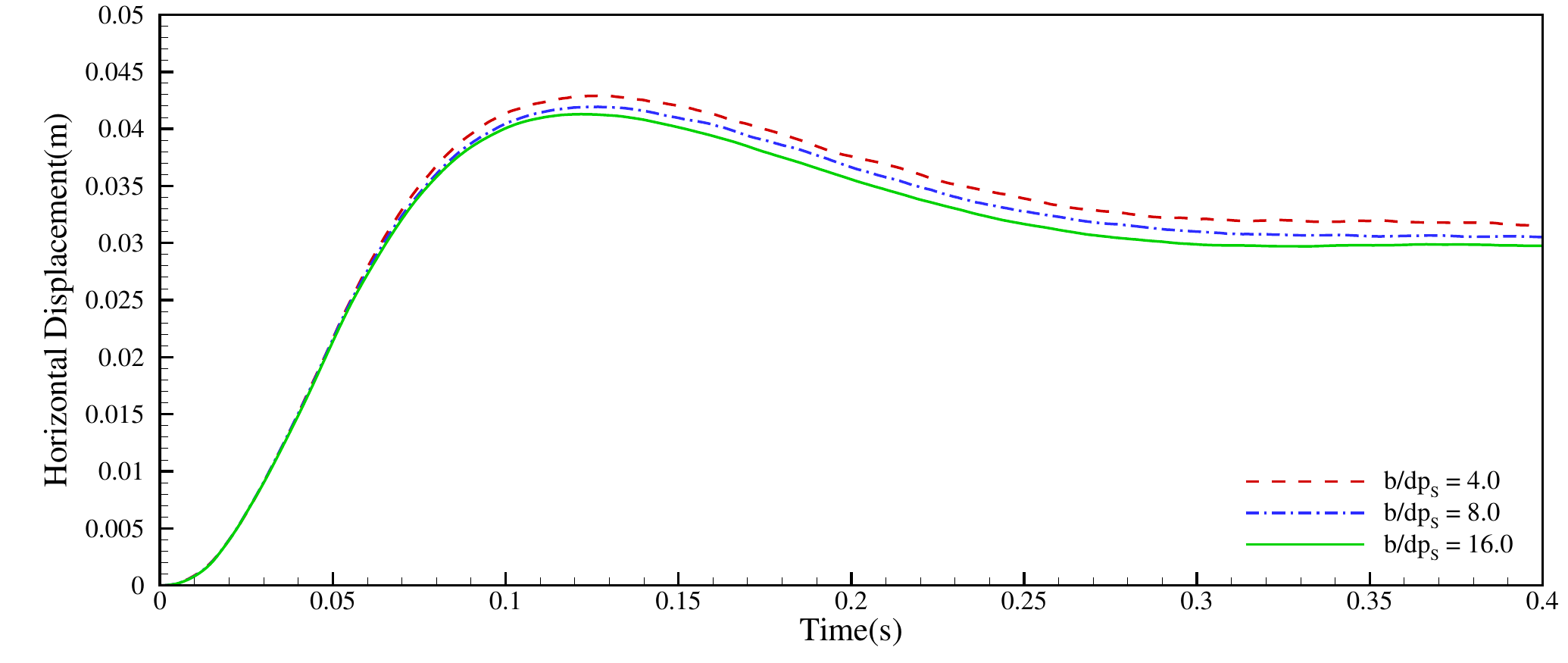}
	\includegraphics[trim = 1mm 2mm 2mm 1mm, clip,width=.85\textwidth]{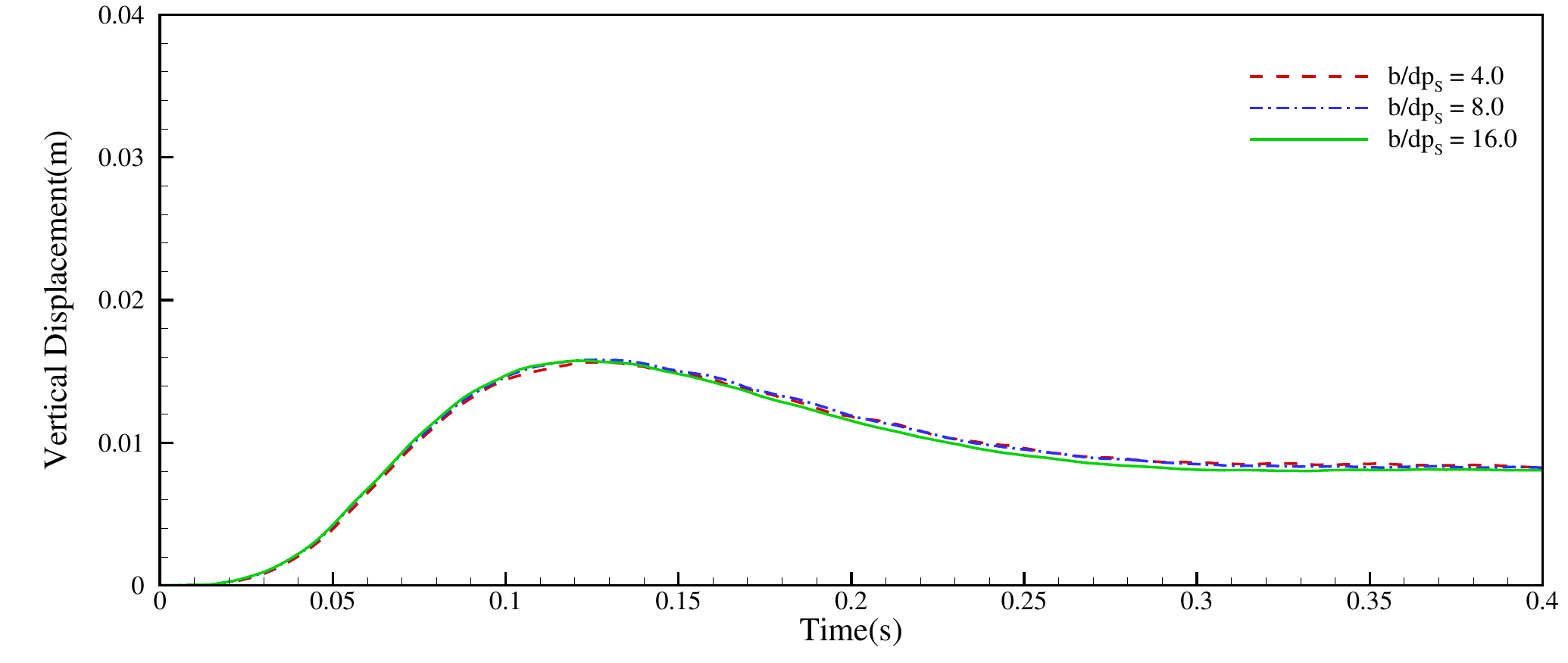}
	\caption{Convergence study for Case-III : 
		horizontal displacement of the free end of the plate (top panel) and  vertical displacement of the free end of the plate (bottom panel).}
	\label{figs:dam-plate-data-convergence}
\end{figure}

The results from a convergence study are plotted in Figure \ref{figs:dam-plate-data-convergence}. 
Before reaching its maximum deformation, 
identical behaviors are noted for three spatial resolutions. 
For the horizontal displacement, 
the maximum deformation decreases slightly as spatial resolution increase, 
which is consistent with the results obtained with single resolution as shown in Figure \ref{figs:dam-plate-data}. 
\begin{table}[htb!]
	\centering
	\caption{Dam-break flow through an elastic gate: computational efficiency. 
		The computer information is similar with that given in Table \ref{tab:fsi-data} and we evaluate the CPU wall-clock time for computation until the time of $0.4$. }
	\begin{tabular}{ccccc}
		Cases      			& 	I & II & III & IV \\
		\hline
		CPU time (s)	&   80.15&523.17 &27.17    &130.77   \\ 
	\end{tabular}
	\label{tab:dam-gate-cpu}
\end{table}
%
\subsection{Pulsatile flow through venous valve leaflets}
\label{subsec:valve}
\begin{figure}[htb!]
	\centering
	\includegraphics[trim = 2.5cm 5cm 2.5cm 5cm, clip,width=0.95\textwidth]{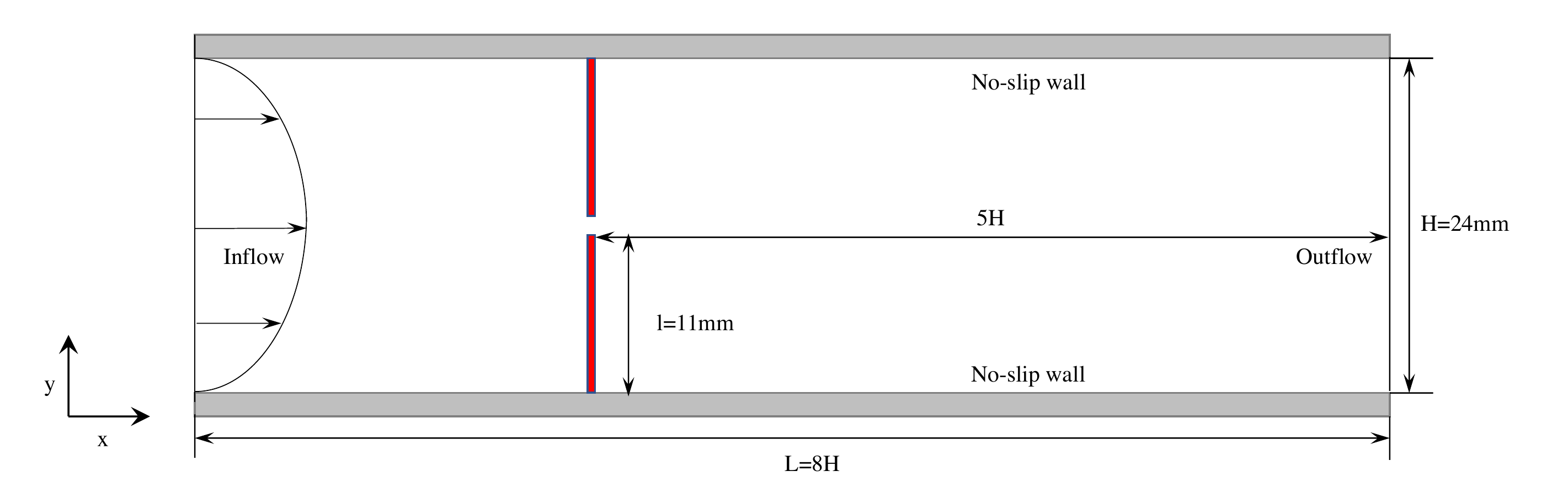}
	\caption{Sketch of the computational setup for the two dimensional idealized vein with flexible valve leaflets.
	}
	\label{figs:valvesetup}
\end{figure}

Here,
we consider a pulsatile flow through a flexible venous valve using a two dimensional idealized geometry 
to demonstrate the versatility of the present method for practical applications in bio-mechanical system.
As depicted in Figure \ref{figs:valvesetup}, 
straight valve leaflets attached to the walls are taken into account and the leaflet thickness is chosen to be $b = 0.2mm$ 
falling in the range measured by van Langevelde et al. \cite{van2010effect} for healthy venous valves. 
Although the present geometric shape is highly simplified, 
the investigation reported herein gives the essential flapping behavior of valve leaflets interacting with pulsatile blood flow.
The channel walls are assumed to be rigid, and no-slip boundary condition is imposed.  
Inflow and outflow conditions are applied at the left and right sides of the domain, respectively. 
The imposed pulsative inflow is characterized by the dimensionless Womersley number, 
$W_o = \frac{H}{2} \sqrt{\frac{\omega}{\eta}}$, 
which represents the ratio of the oscillatory inertial effects to the viscous effects, 
where $\omega$ denotes the angular frequency.
The velocity profile of the pulsatile blood-like flow is given in Appendix B. 
The fluid density $\rho_F = 1000 kg/m^3$ and the dynamic viscosity varies according to the Womersley number.
The Neo-Hookean material model is used for the leaflets with Young's modulus $E = 1.5 MPa$, 
Possion's ratio $\nu = 0.49$ and density $\rho_S = 1060 kg/m^3 $.
The initial particle spacing of valve leaflet is set as ${dp}_S = b/4.0$ 
and the resolution ratio is ${dp}_F/{dp}_S = 2.0$.

\begin{figure}[htb!]
	\centering
	\includegraphics[trim = 1mm 1mm 1mm 1mm, clip,width=.95\textwidth]{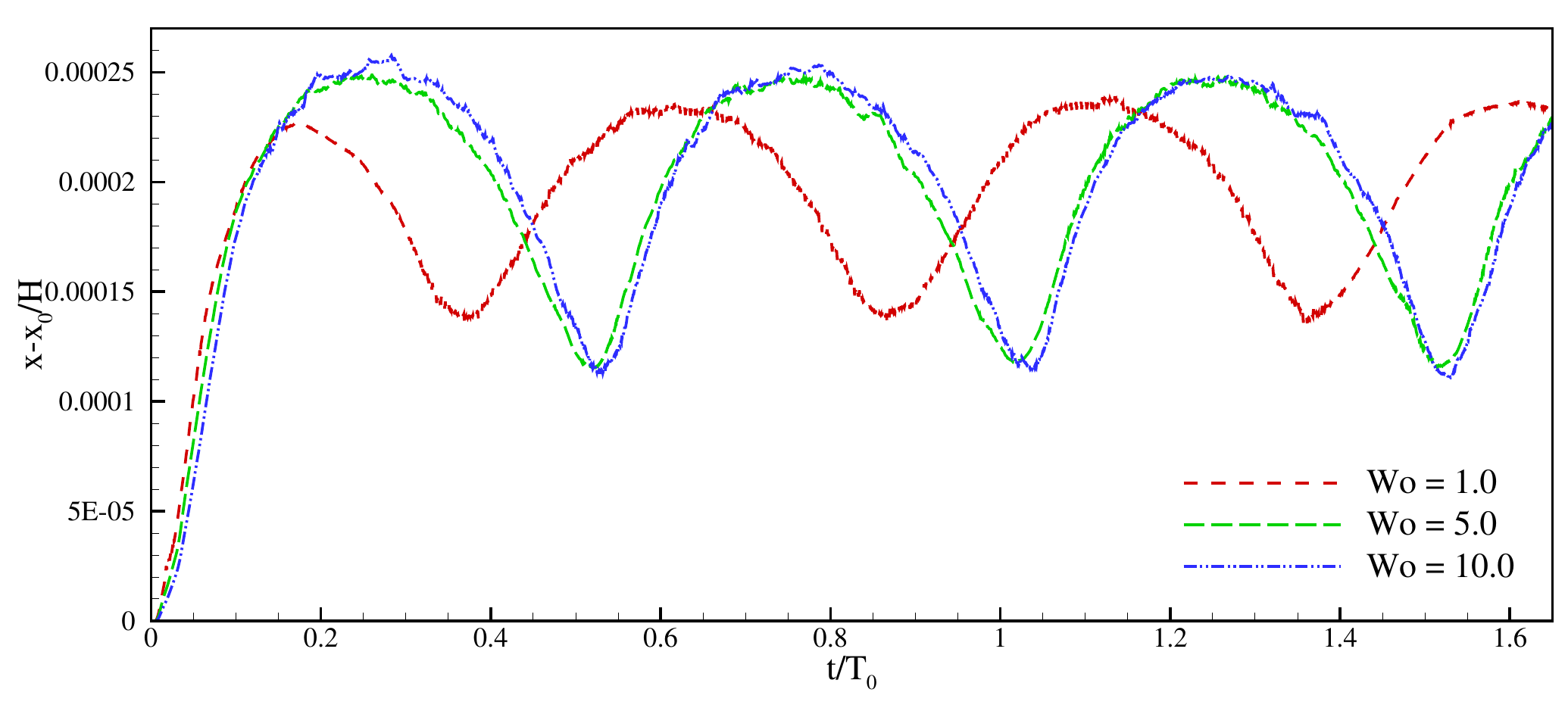}
	\includegraphics[trim = 1mm 1mm 1mm 1mm, clip,width=.95\textwidth]{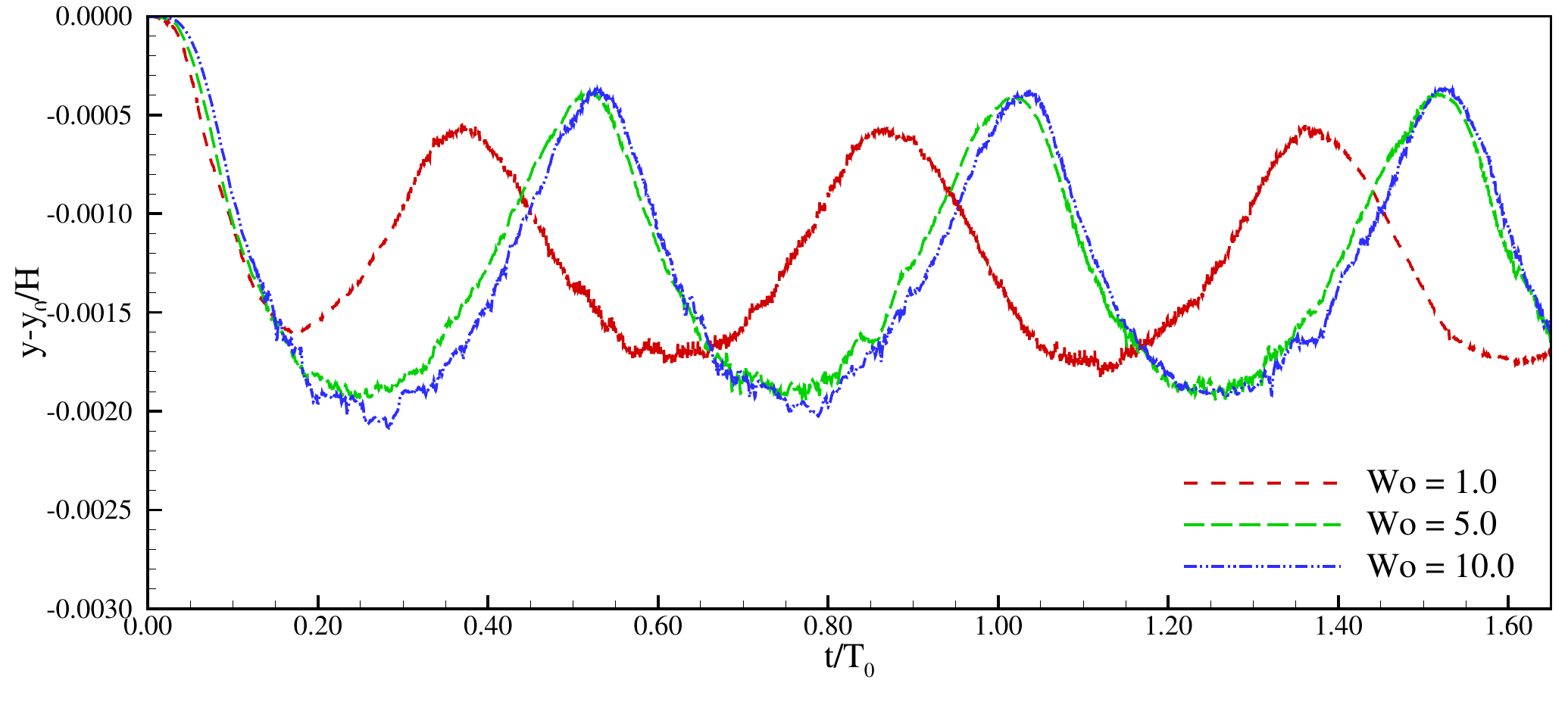}
	\caption{Pulsatile flow through venous valve leaflets:
		horizontal displacement (top panel) and vertical displacement (bottom panel) of the free-end of the top leaflet.
		Computational results for three Womersley numbers are presented. }
	\label{figs:valve-data}
\end{figure}

\begin{figure}[htb!]
	\centering
	\includegraphics[trim = 1mm 16mm 1mm 16mm, clip,width=.95\textwidth]{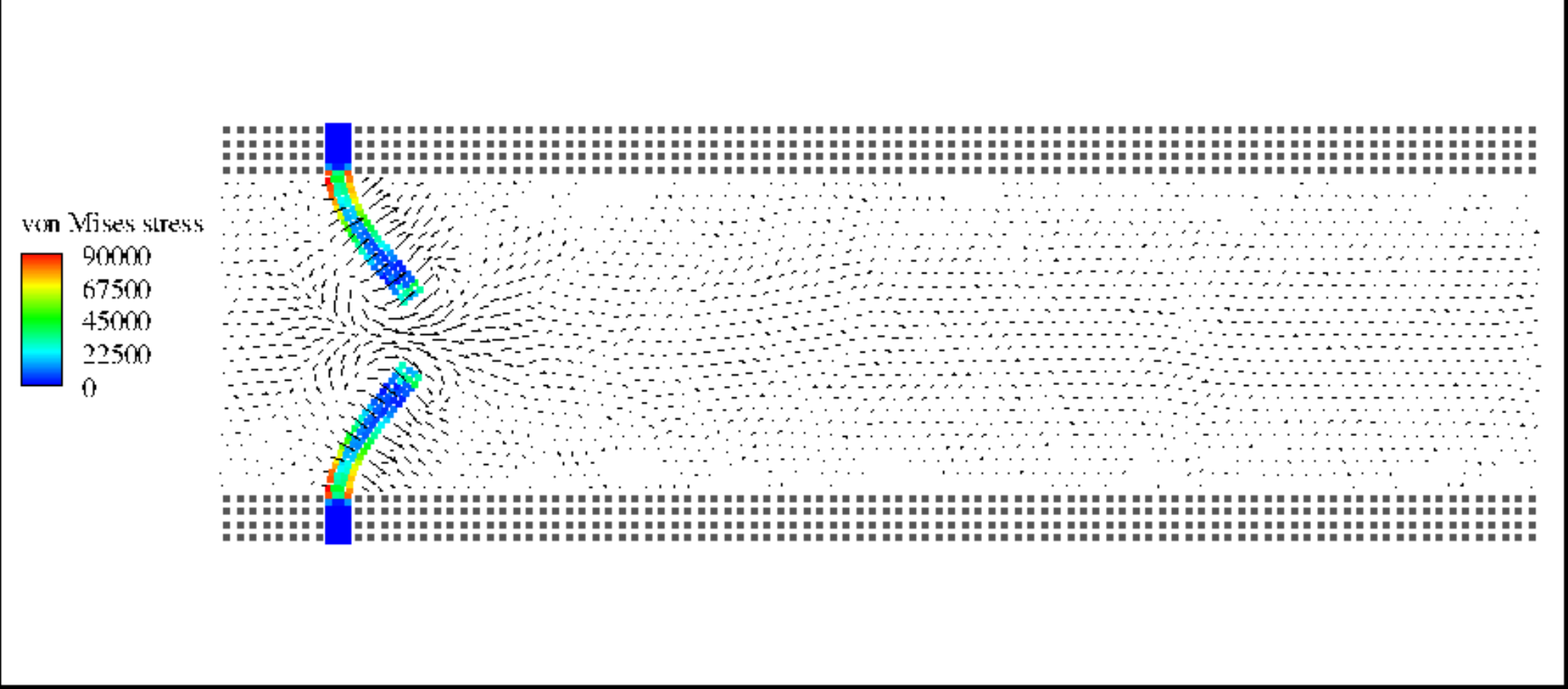}
	\includegraphics[trim = 1mm 16mm 1mm 16mm, clip,width=.95\textwidth]{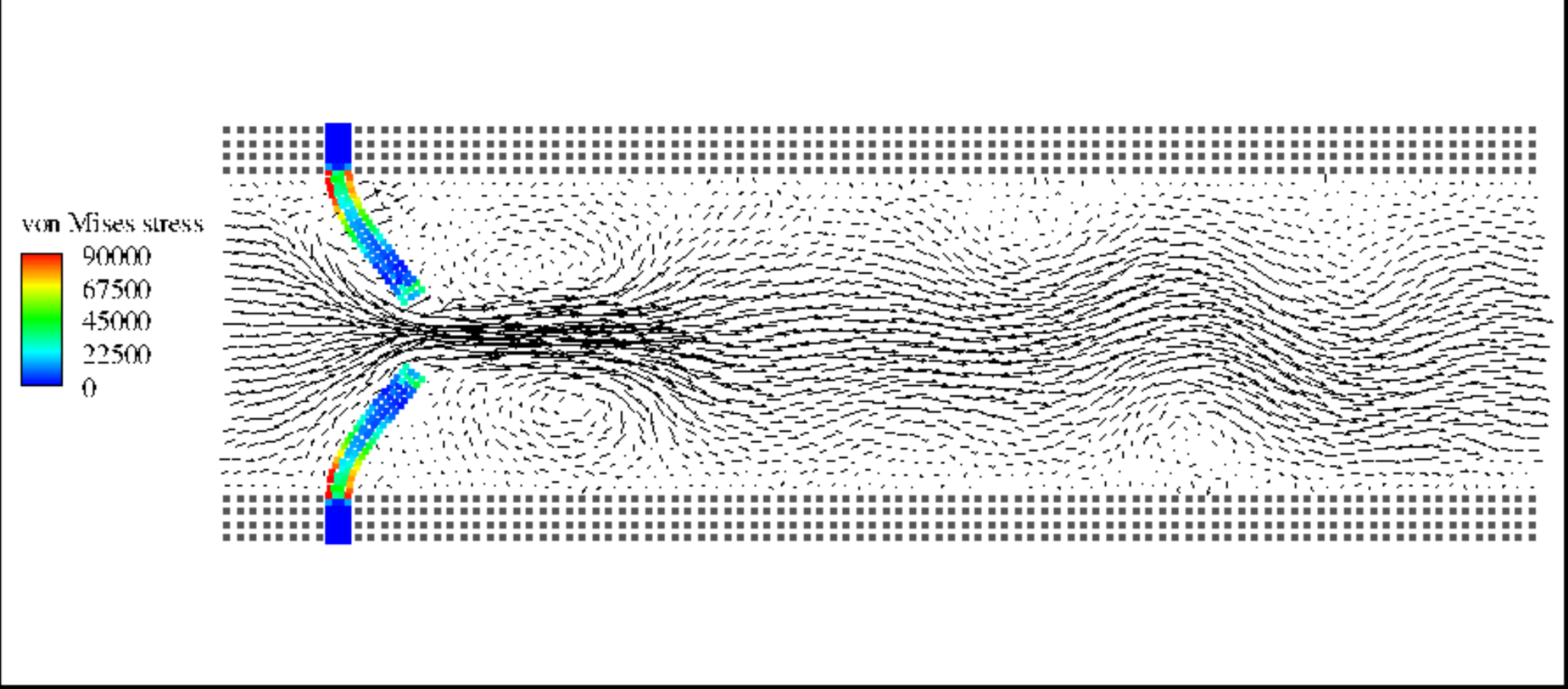}
	\includegraphics[trim = 1mm 16mm 1mm 16mm, clip,width=.95\textwidth]{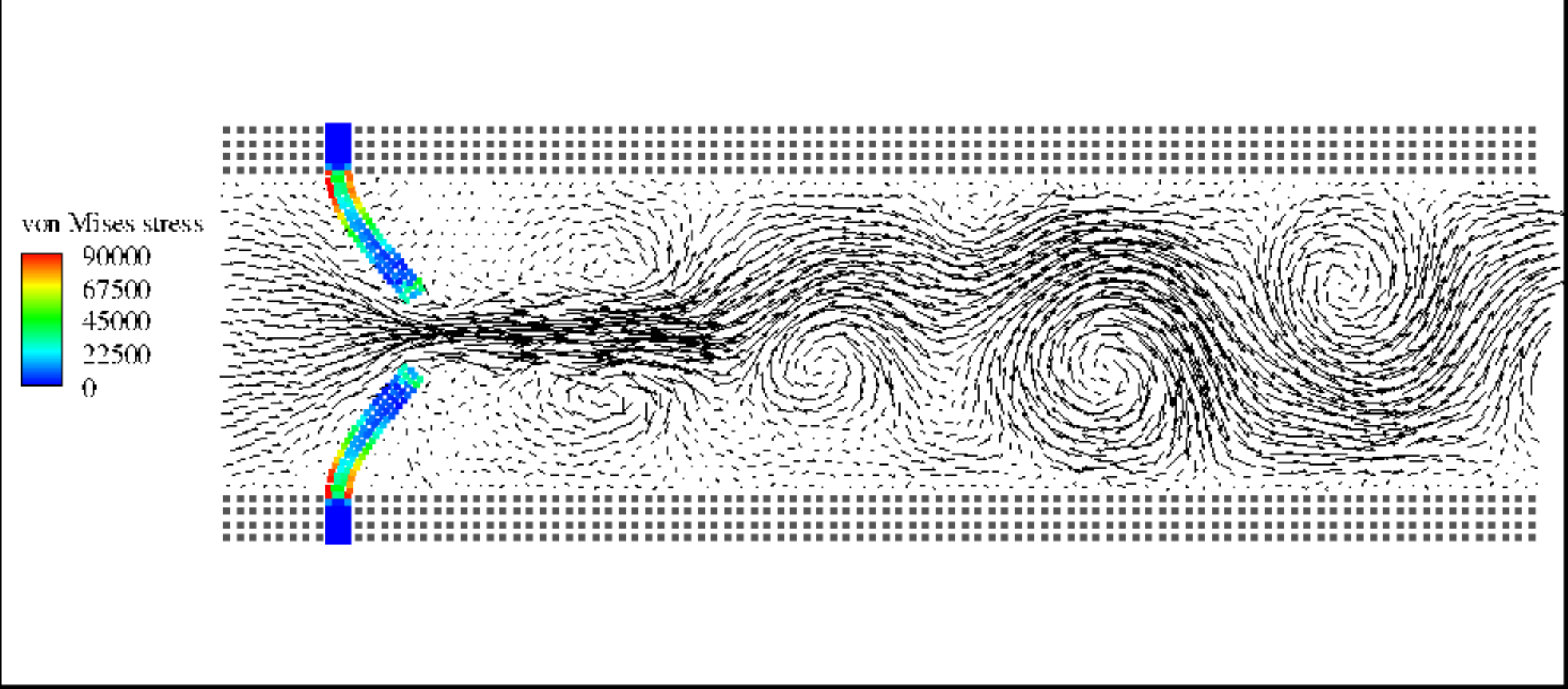}
	\caption{Pulsatile flow through venous valve leaflets: 
		deformation of leaflets colored by von Mieses stress and velocity vectors 
		of the flow.
		Computational results for three Womersley numbers are presented, 
		viz. $W_o = 1.0$ (top panel), $W_o = 5.0$ (middle panel) and $W_o = 10.0$ (bottom panel).}
	\label{figs:valve-com}
\end{figure}

\begin{figure}[htb!]
	\centering
	\includegraphics[trim = 10mm 15mm 1mm 20mm, clip,width=.95\textwidth]{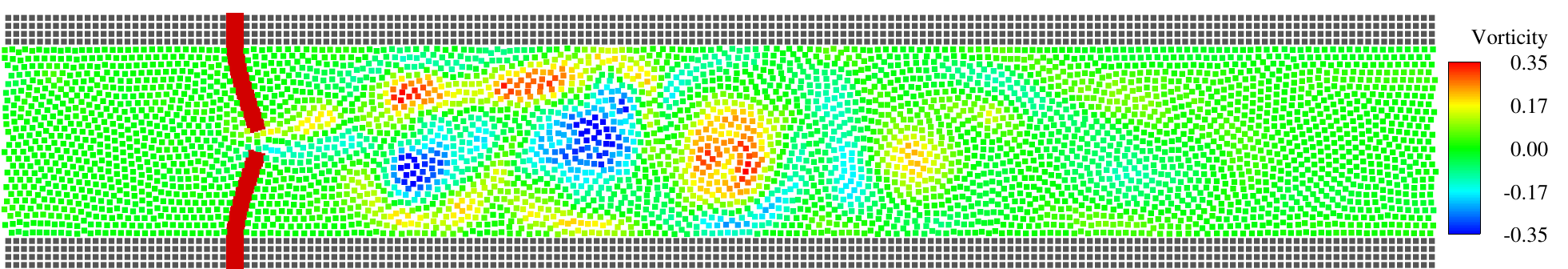}
	\includegraphics[trim = 10mm 15mm 1mm 20mm, clip,width=.95\textwidth]{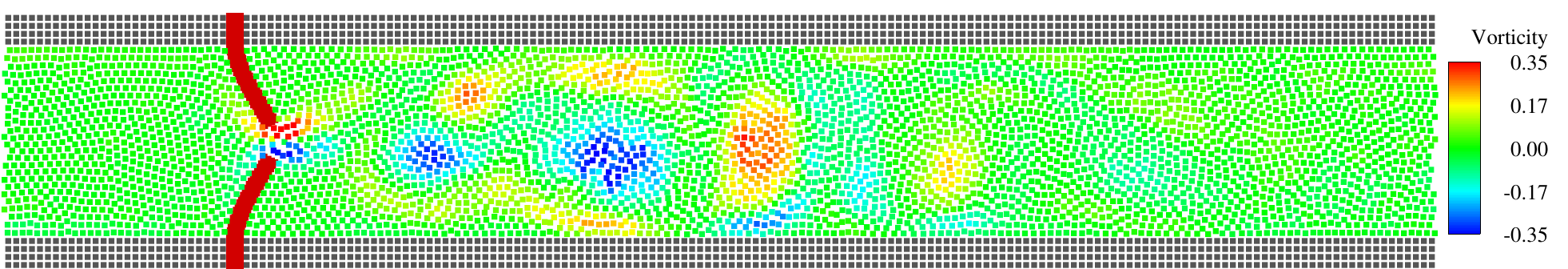}
	\includegraphics[trim = 10mm 15mm 1mm 20mm, clip,width=.95\textwidth]{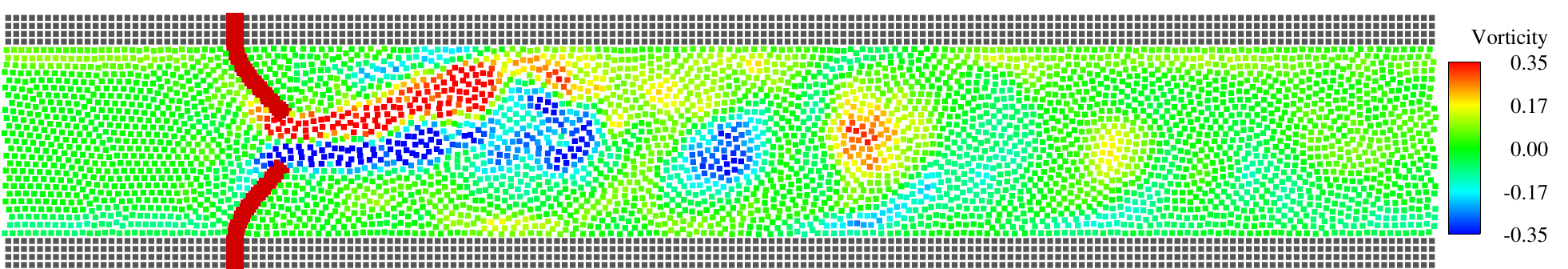}
	\includegraphics[trim = 10mm 15mm 1mm 20mm, clip,width=.95\textwidth]{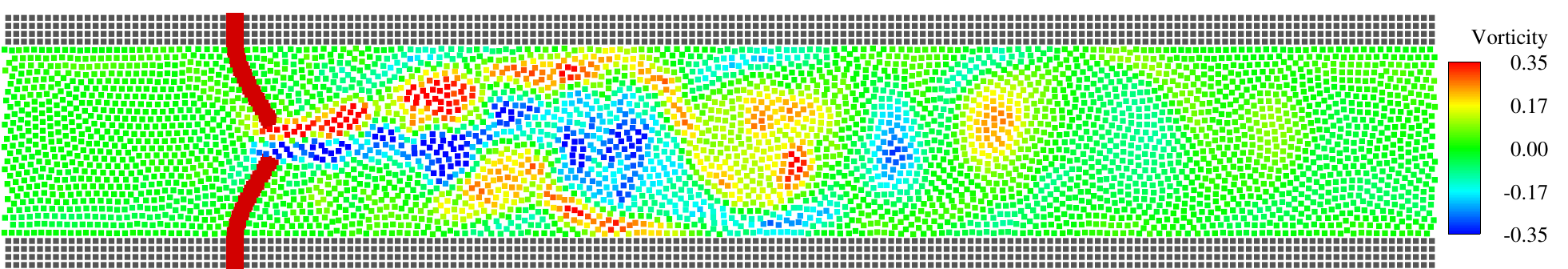}
	\caption{Pulsatile flow through venous valve leaflets: 
		deformation of leaflets and the corresponding vorticity field at different time instants,
		viz. $t = 0.62~s$ (first panel), $t=0.65~s$ (second panel), $t=0.75~s$ (third panel) and $t=0.86~s$(last panel).
		The Womersley number is $W_o = 10.0$.
	}
	\label{figs:valve-wo-10}
\end{figure}

We consider three cases with different Womersley numbers, 
namely, $W_o = 1.0, ~5.0$ and $10.0$, corresponding to the velocity profiles shown in Figure \ref{figs:pulsatile-v}. 
Figure \ref{figs:valve-com} gives the deformation of leaflets colored by the von Mises 
stress and the corresponding velocity vectors at $t = 2.0~s$.
A series of vortices in the downstream of the leaflets can be observed.  
They are particularly prevalent when the Womersley number is large. 
Figure \ref{figs:valve-wo-10} depicts several frames of the flow field colored by 
vorticity and the deformation of the valve leaflets with $W_o = 10.0$.
Four snapshots approximately span a flapping period. 
Large velocity gradients develop at the tip of the leaflet and a shear 
layer originates from the tip and separates the recirculating flow region 
behind the leaflet from the main flow.
The predicted horizontal and vertical displacements of the top leaflet are given in Figure \ref{figs:valve-data}. 
It is observed that the interactions between the pulsatile flow and the leaflets 
give rise to self-sustained large-amplitude oscillations with the similar time period of the pulsatile inflow. 
With increasing Womersley number, 
large-amplitude oscillations are observed while the period remains unaltered from that of the pulsatile inflow
%
\subsection{Passive flapping of flexible fish-like body}\label{test-fish}
\begin{figure}[htb!]
	\centering
	\includegraphics[trim = 2.5cm 4cm 2.5cm 2cm, clip,width=0.95\textwidth]{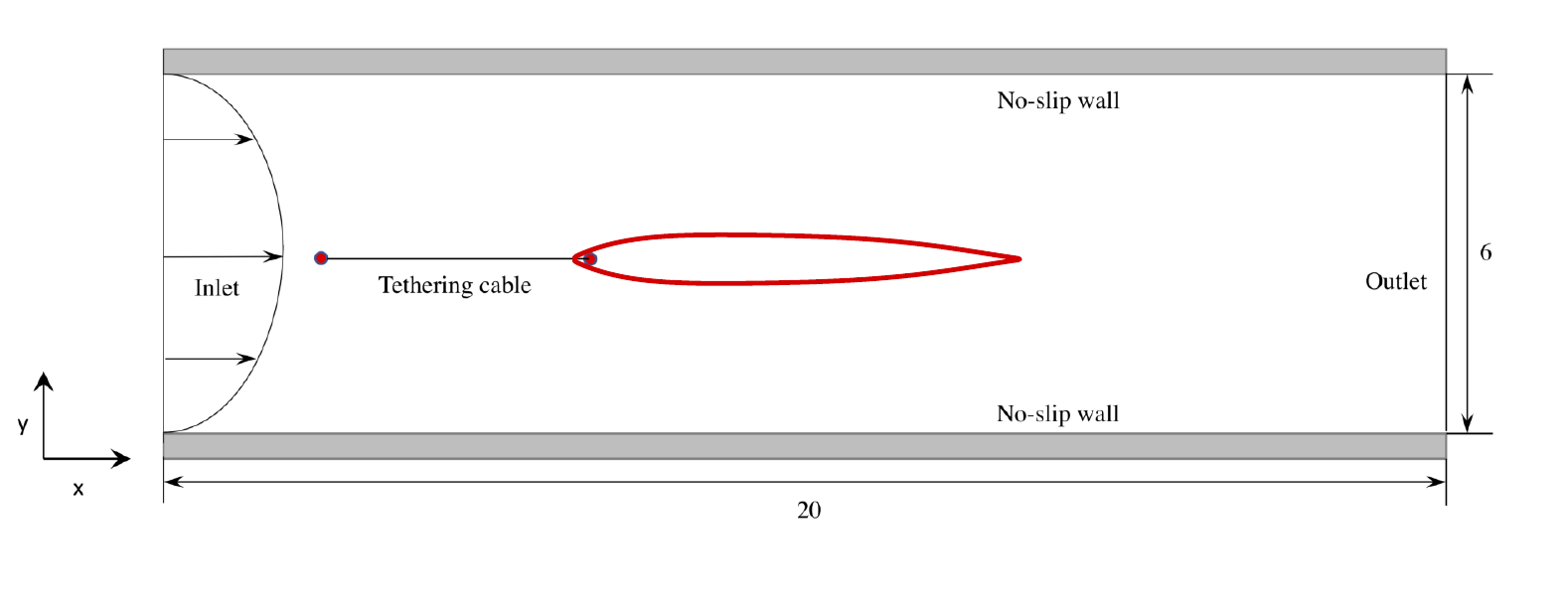}\\
	\includegraphics[trim = 0cm 2cm 0cm 2.5cm, clip,width=0.65\textwidth]{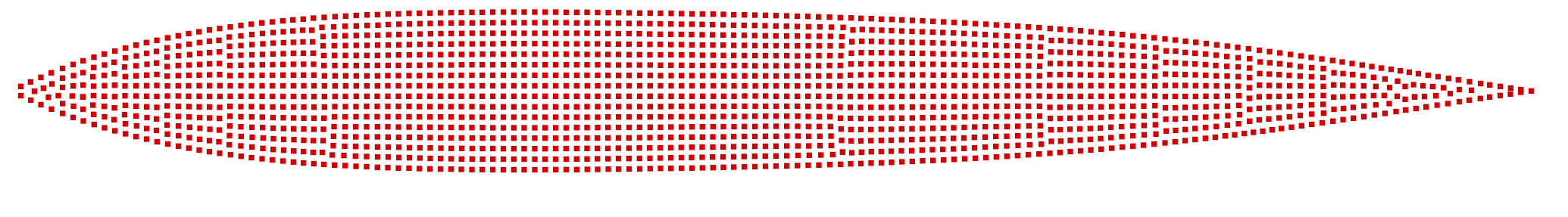}
	\caption{Schematic diagram of the two dimensional flow induced passive flapping of the flexible fish-like body, and the corresponding initial particle distribution of the body.}
	\label{figs:fishsetup}
\end{figure}

Here, 
the passive flapping of a flexible fish-like body is investigated to validate the versatility of the present method.
Different from previous simulations involving passive flapping of flexible filaments (see e.g. \cite{huang2007simulation}),
the present fish-like body has a foil-like shape.
The body is free except its head is tethered to a fix point, 
located at (6.0, 3.0), through a $4.2$ long tethering cable.
The schematics of the computational setup and the initial particle distribution of the body are shown in Figure \ref{figs:fishsetup}.
The outline of the body is given by a mirrored $5^{th}$ order polynomial function \cite{curatolo2016modeling} 
$y = c(x)  = a_i x^i, i= 1,2,...5$, with coefficients $a_1 = 1.22 bh / L$, $a_2 = 3.19 bh / L^2$, 
$a_3 = -15.73 bh / L^3$, $a_4 = 21.87 bh / L^4$ and $a_5 = -10.55 bh / L^5$,
where $bh = 0.4$ represents the maximum body thickness and $L = 3.738$ denotes the body length.
No-slip boundary condition is imposed on the top and bottom walls, 
while inflow and outflow conditions are applied at the left and right sides of the domain, respectively. 
The inflow is realized by a velocity profile of $U(y) = 1.5 \overline{U}(t) y (H-y) / H^2$ with a smooth procedure
$ \overline{U}(t) =0.5(1.0 - \cos(0.5 \pi t))$ if $ t \leq 4.0$ otherwise  $ \overline{U}(t) =1.0$.
The fluid density $\rho_F = 1.0 $ and the Reynolds number $Re = \frac{U_0 L}{\eta} = 5000$.
The Neo-Hookean material model is used for the body 
with dimensionless Young's modulus $E^*= E/{\rho_F U_0^2} = 500$, 
Possion's ratio $\nu = 0.49$ and density $\rho_S = \rho_F $. 
To discretize the system, 
the initial particle spacing of the body is about ${dp}_S = bh/8.0$ and the resolution ratio ${dp}_F/{dp}_S = 2.0$.

\begin{figure}[htb!]
	\centering
	\includegraphics[trim = 1mm 10mm 1mm 12mm, clip,width=.85\textwidth]{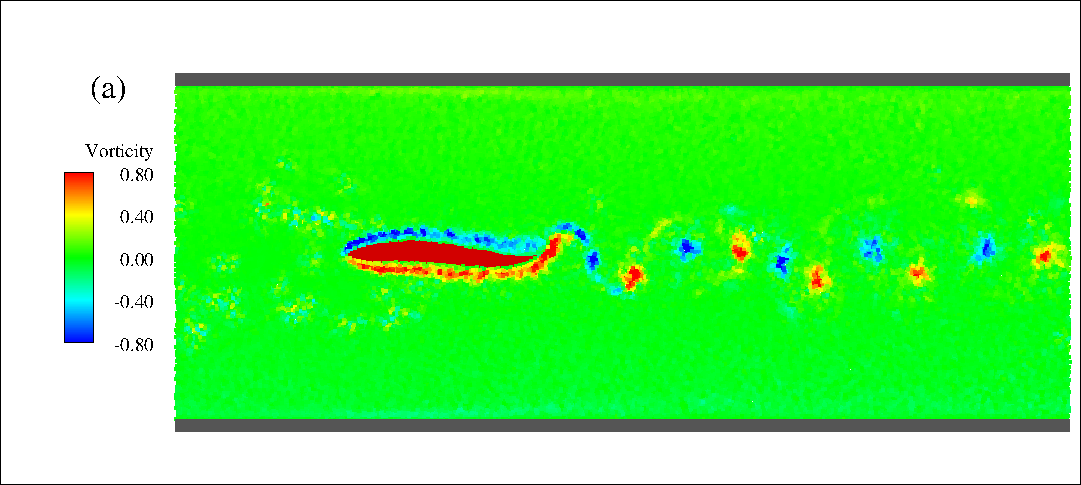}\\
	\includegraphics[trim = 1mm 10mm 1mm 12mm, clip,width=.85\textwidth]{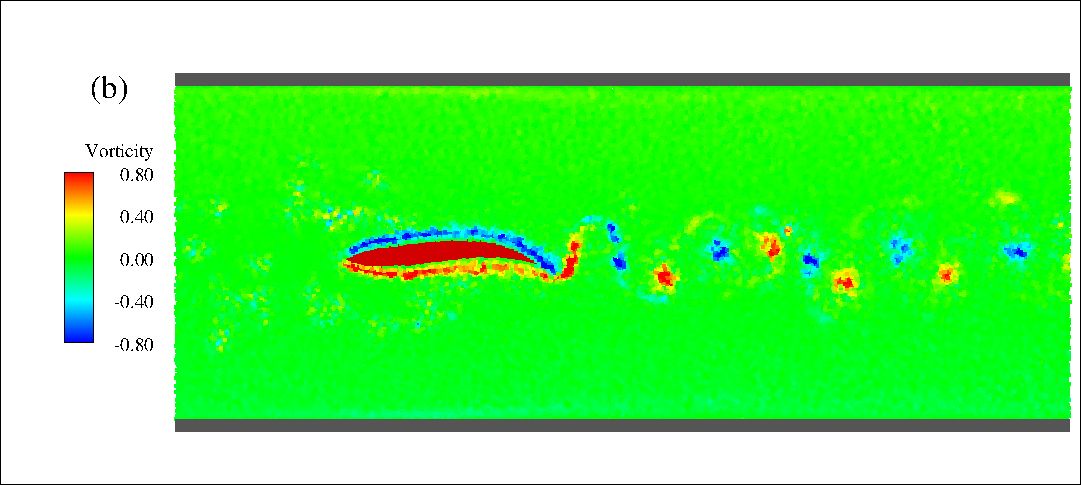}\\
	\includegraphics[trim = 1mm 10mm 1mm 12mm, clip,width=.85\textwidth]{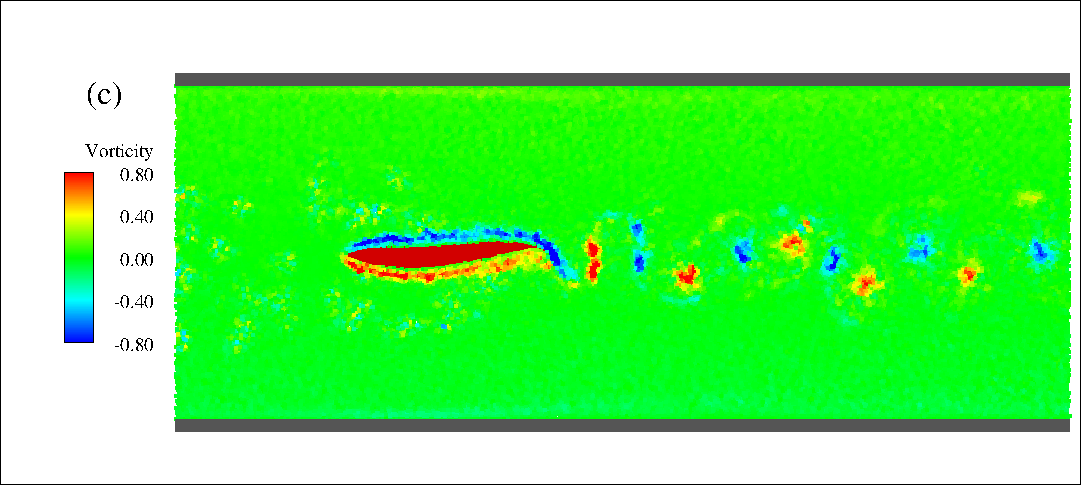}\\
	\includegraphics[trim = 1mm 10mm 1mm 12mm, clip,width=.85\textwidth]{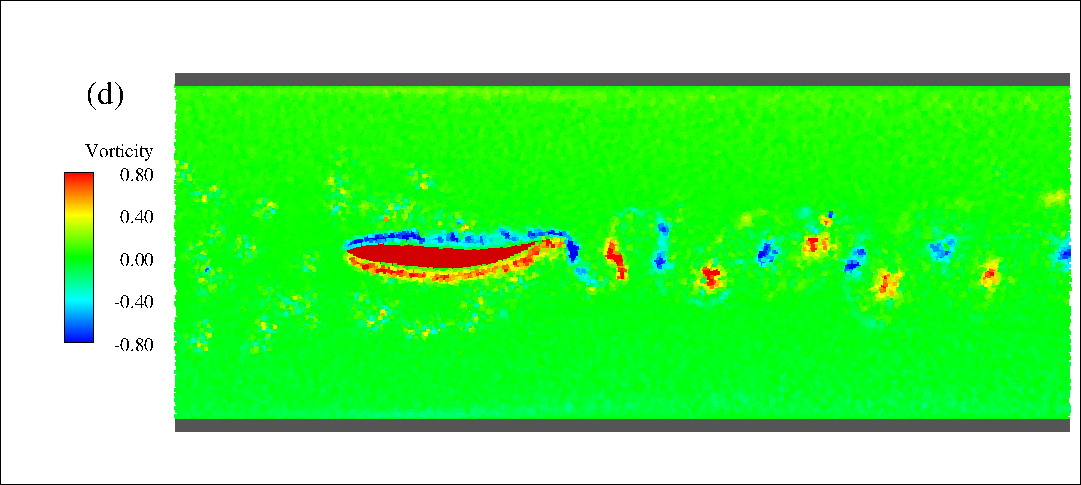}
	\caption{Passive flapping of the flexible fish-like body: snapshot of the body deformation 
		and the vorticity around the free swimmer at various time instants, viz. 
		(a) $t = 48.8$, (b) $t = 49.3$, (c) $49.6$ and (d) $49.9$.}
	\label{figs:fish-vorticity}
\end{figure}

Figure \ref{figs:fish-vorticity} depicts snapshots of the flow field colored by 
the vorticity and the body deformation.
The four frames approximately span a passive flapping period.
As the passive flapping has reached a sustained stage, 
the distribution of vorticity forms a sheet line behind the body and
two large vortices are shed in each passive flapping period from the tail. 
Vortices opposite signs are shed alternately from the body at the moment when it is most bent.
The trend in the flow pattern is in agreement with the experimental results of Zhang et al. \cite{zhang2000flexible}, 
which showed that a sheet of vortices is produced by a flexible filaments in a flowing soap file.

\begin{figure}[htb!]
	\centering
	\includegraphics[trim = 2mm 2mm 2mm 2mm, clip,width=.5\textwidth]{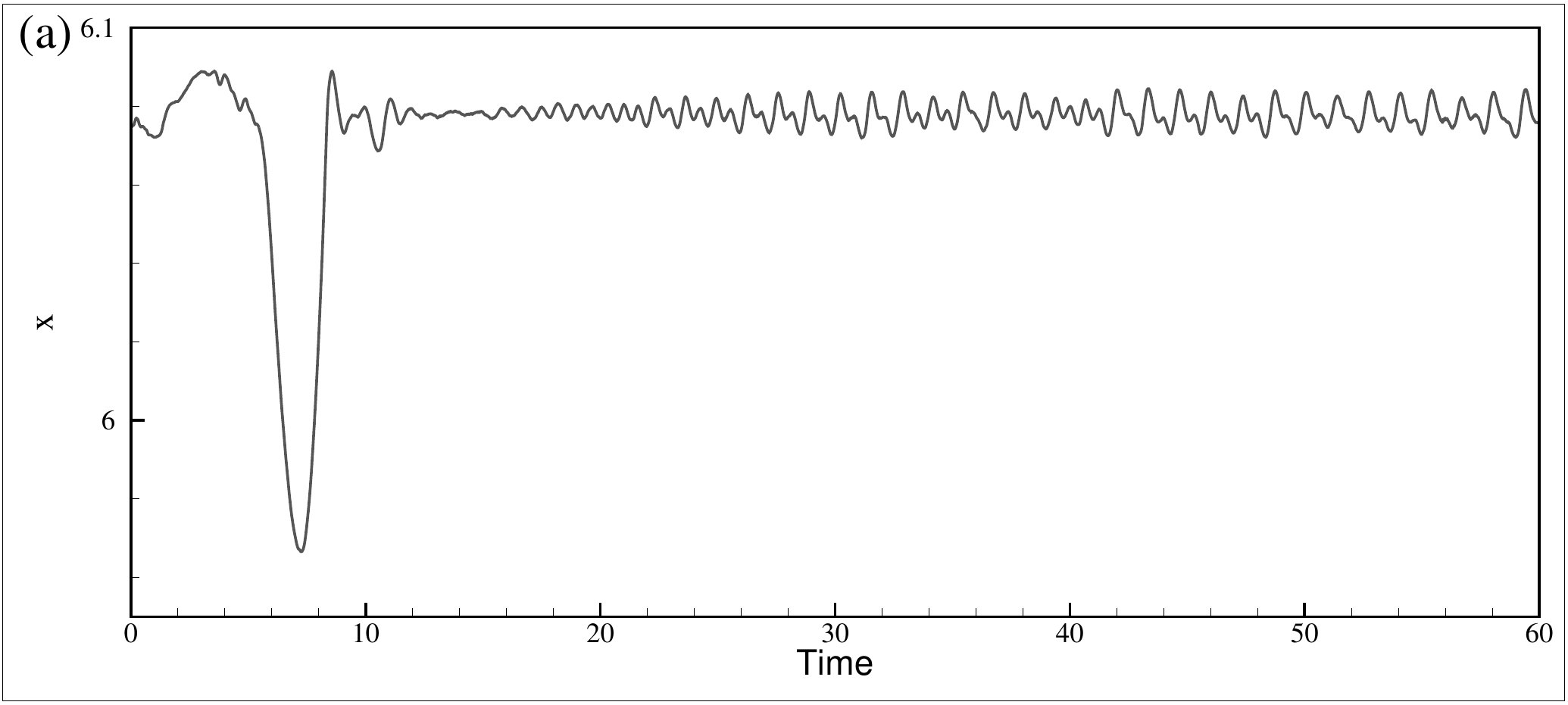}\\
	\includegraphics[trim = 2mm 2mm 2mm 2mm, clip,width=.5\textwidth]{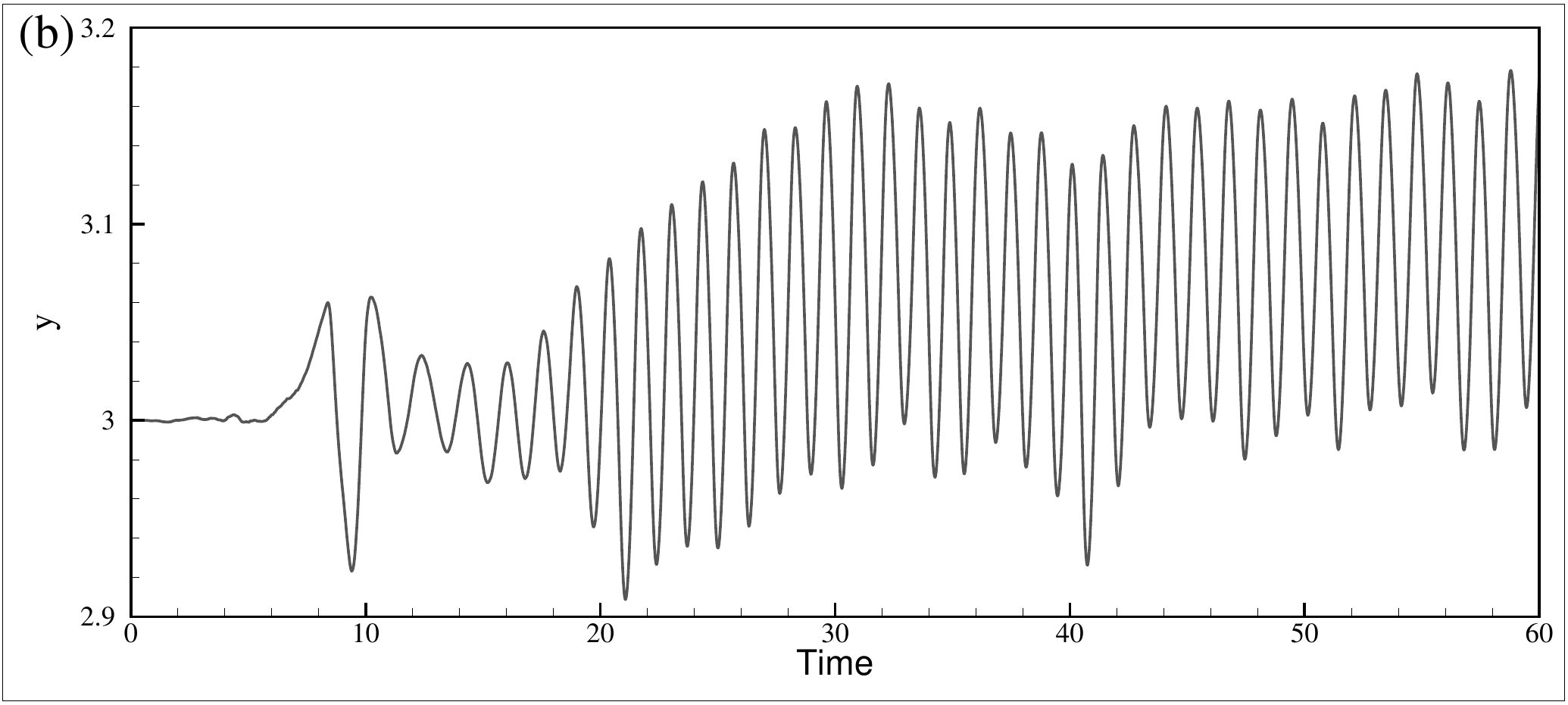}\\
	\includegraphics[trim = 2mm 2mm 2mm 2mm, clip,width=.5\textwidth]{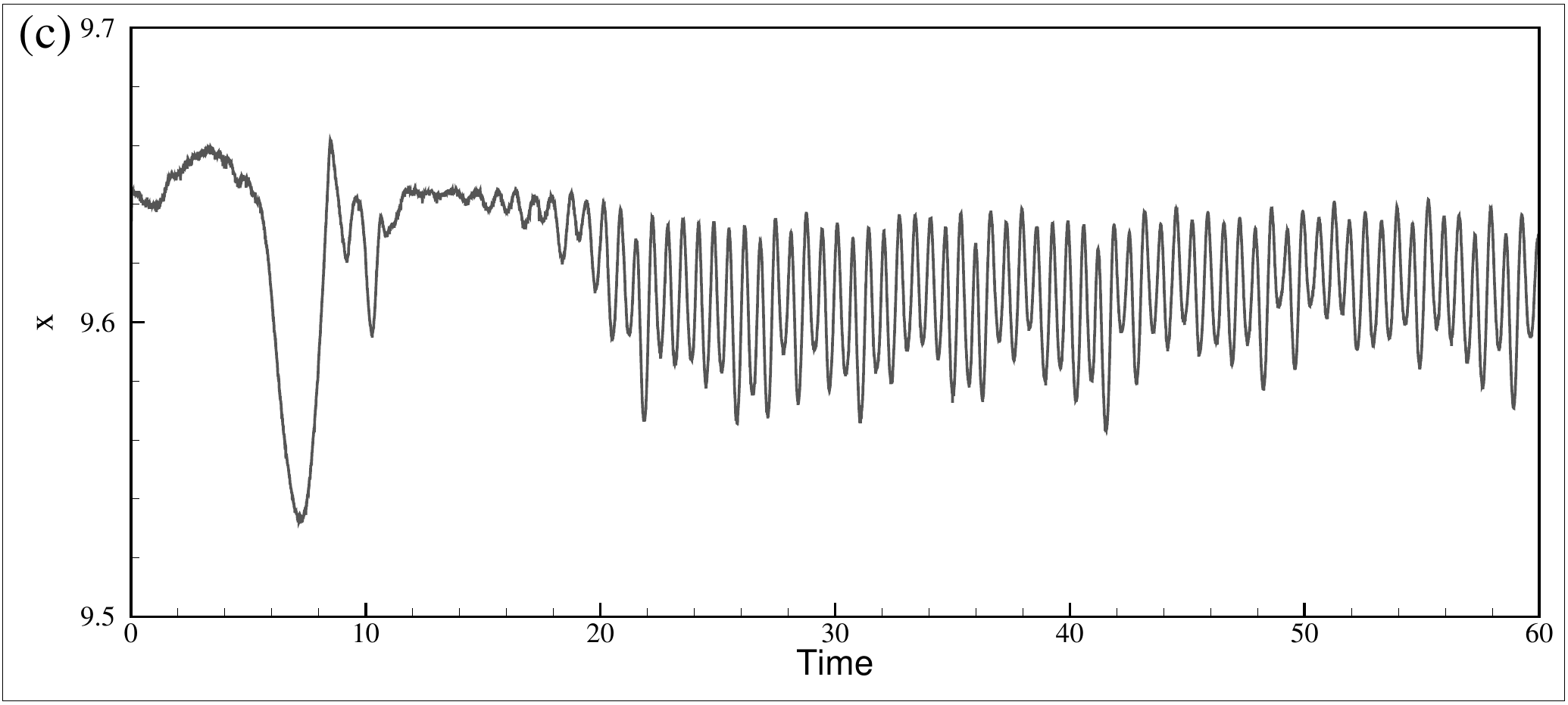}\\
	\includegraphics[trim = 2mm 2mm 2mm 2mm, clip,width=.5\textwidth]{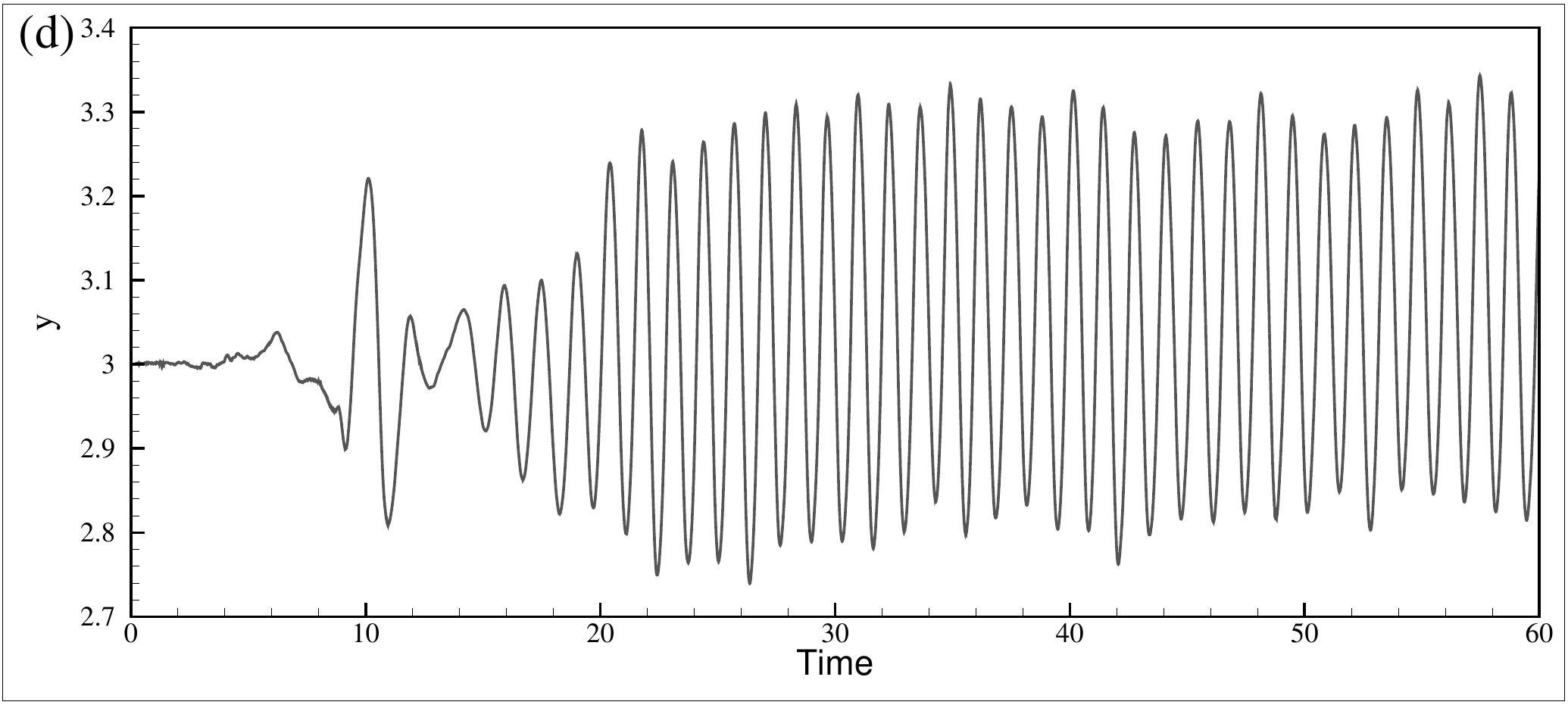}\\
	\includegraphics[trim = 2mm 2mm 2mm 2mm, clip,width=.5\textwidth]{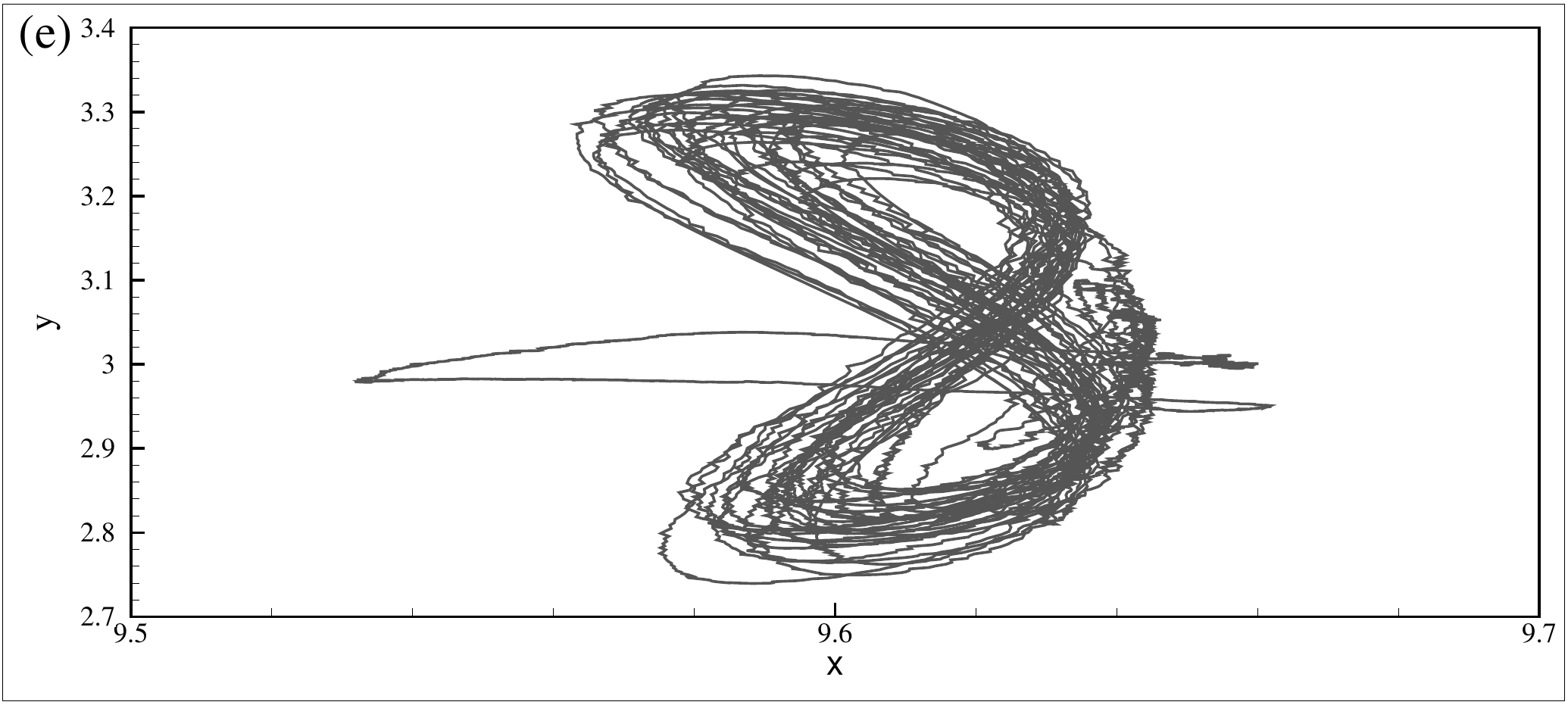}
	\caption{Time histories of the head and tail positions of a flexible fish-like body: 
		(a) head position in $x$-direction,
		(b) head position in $y$-direction,
		(c) tail position in $x$-direction,
		(d) tail position in $y$-direction, and
		(e) the trajectory of the tail.} 
	\label{figs:fish-data}
\end{figure}

\begin{figure}[htb!]
	\centering
	\includegraphics[trim = 5mm 2.75cm 4mm 2.5cm, clip,width=\textwidth]{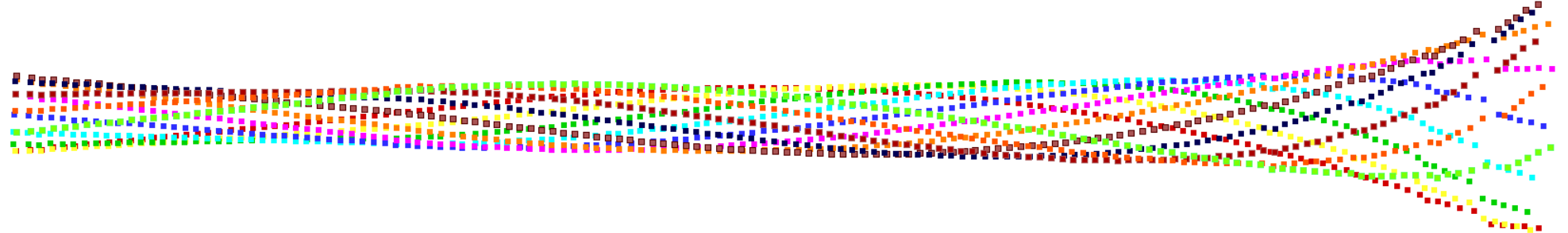}
	\caption{Flapping pattern of the flexible fish-like body. Particles located at the central line of the body for a typical period of flapping.} 
	\label{figs:fish-center-particle}
\end{figure}

In Figure \ref{figs:fish-data}, 
the $x$ and $y$ positions of the head and tail are plotted as functions of time.
Sustained flapping is achieved after non-dimentional time $t = 20$.
It is also noted that the preliminary passive flapping frequency of the tail is around $0.74 s^{-1}$. 
Furthermore, the approximated vertical amplitude of head and tail are $0.3$ and $0.58$, 
i.e. $37.5\%$ and $72.5\%$ of the size of the body width, respectively.
Such large deformations justify the robustness of the present method in modeling applications in bio-mechanical system. 
It is worth noting that the trajectory of the tail shows a typical Lissajous curve with a frequency ratio about $2.25:1$ between the horizontal and vertical components.
The flexible body undergoes a traveling wave motion as shown in Figure \ref{figs:fish-center-particle}, 
which is similar to the typical flapping pattern of an active fish swimming.
%
%
%
\section{Concluding remarks}\label{sec:conclusion}
In this paper, 
we have presented a simple, efficient and accurate multi-resolution SPH method 
for fluid-structure interactions.
The present method is capable of incorporating different spatio-temporal resolutions 
for the fluid and structure, 
and demonstrates a considerably improved computational efficiency.
A position-based Verlet scheme is introduced to ensure the momentum conservation in fluid-structure coupling 
when multiple time steps are applied. 
Furthermore, 
the time-averaged velocity and acceleration of solid particles are introduced to enhance the force matching in fluid-structure coupling. 
Extensive numerical examples demonstrate the robustness and reasonable accuracy in reproducing the solid dynamics when a coarser resolution is sufficient for the fluid dynamics. 
Special attention is paid to test the present method 
for modeling applications in bio-mechanical systems, 
e.g. venous valve and passive flapping of fish-like body. 
Though the model used here is quite simple, 
the resultant good performance shows promising potential to future more practical applications. 
%
%
\section{Acknowledgement}
The authors would like to express their gratitude to Deutsche Forschungsgemeinschaft for their sponsorship of this research under grant number DFG HU1572/10-1 and DFG Hu1527/12-1.
%
%
\section*{References}
\bibliography{mybibfile}
%
%
\section*{Appendices}
\subsection*{Appendix A : Velocity profile of the pulsative flow}
\begin{figure}[H]
	\centering
	\includegraphics[trim = 0.5cm 0cm 1.5cm 0cm, clip,width=.3\textwidth]{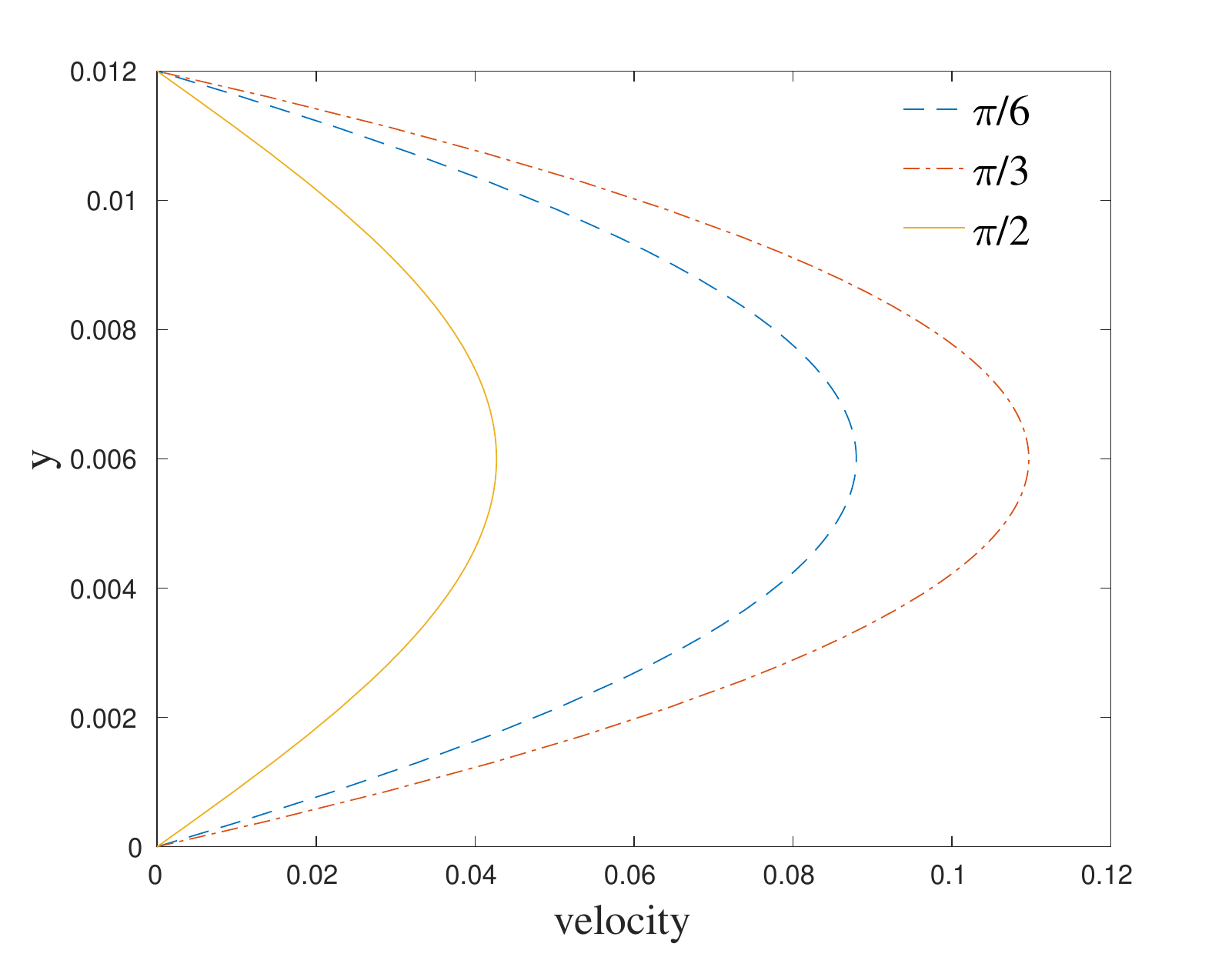}
	\includegraphics[trim = 0.5cm 0cm 1.5cm 0cm, clip,width=.3\textwidth]{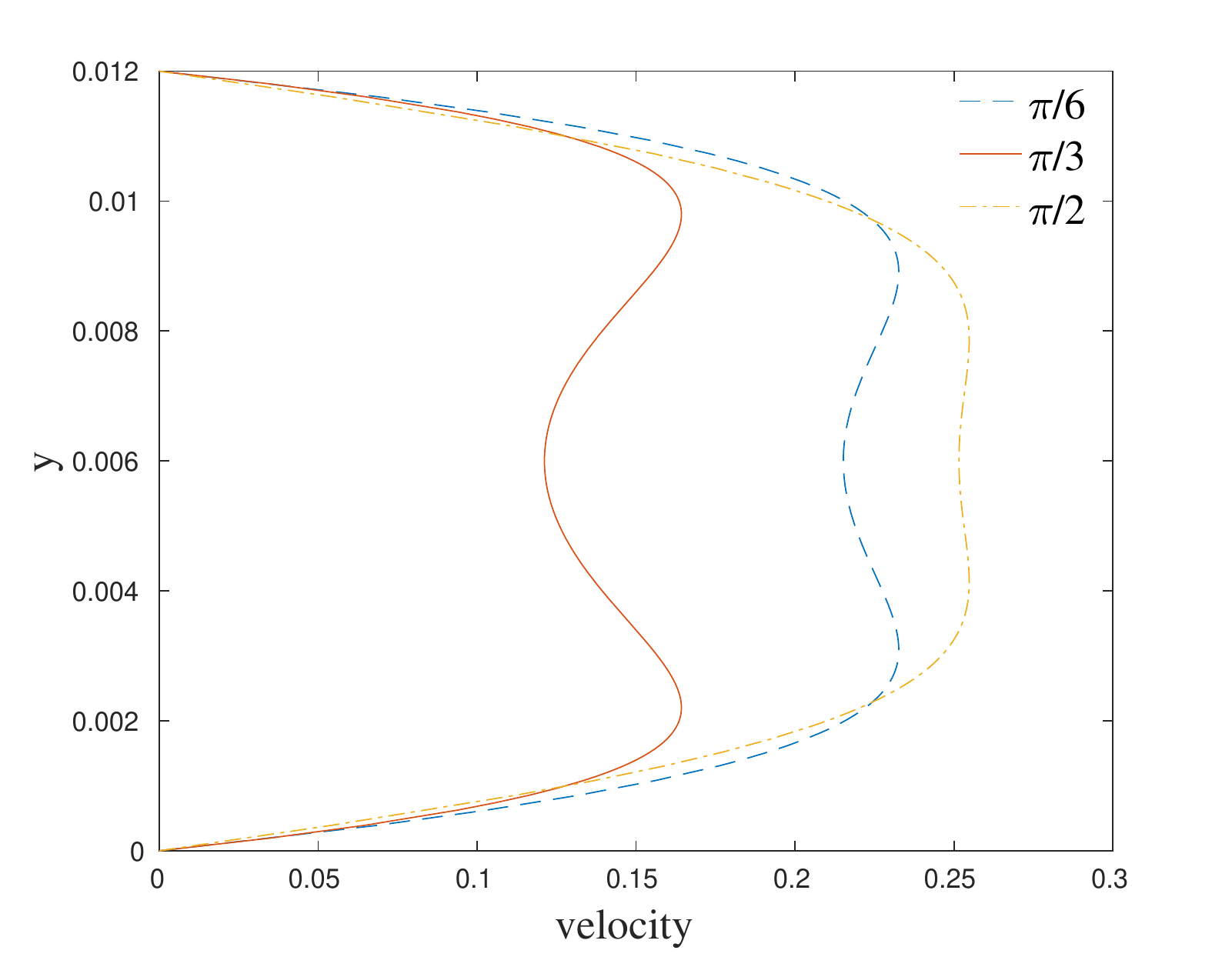}
	\includegraphics[trim = 0.5cm 0cm 1.5cm 0cm, clip,width=.3\textwidth]{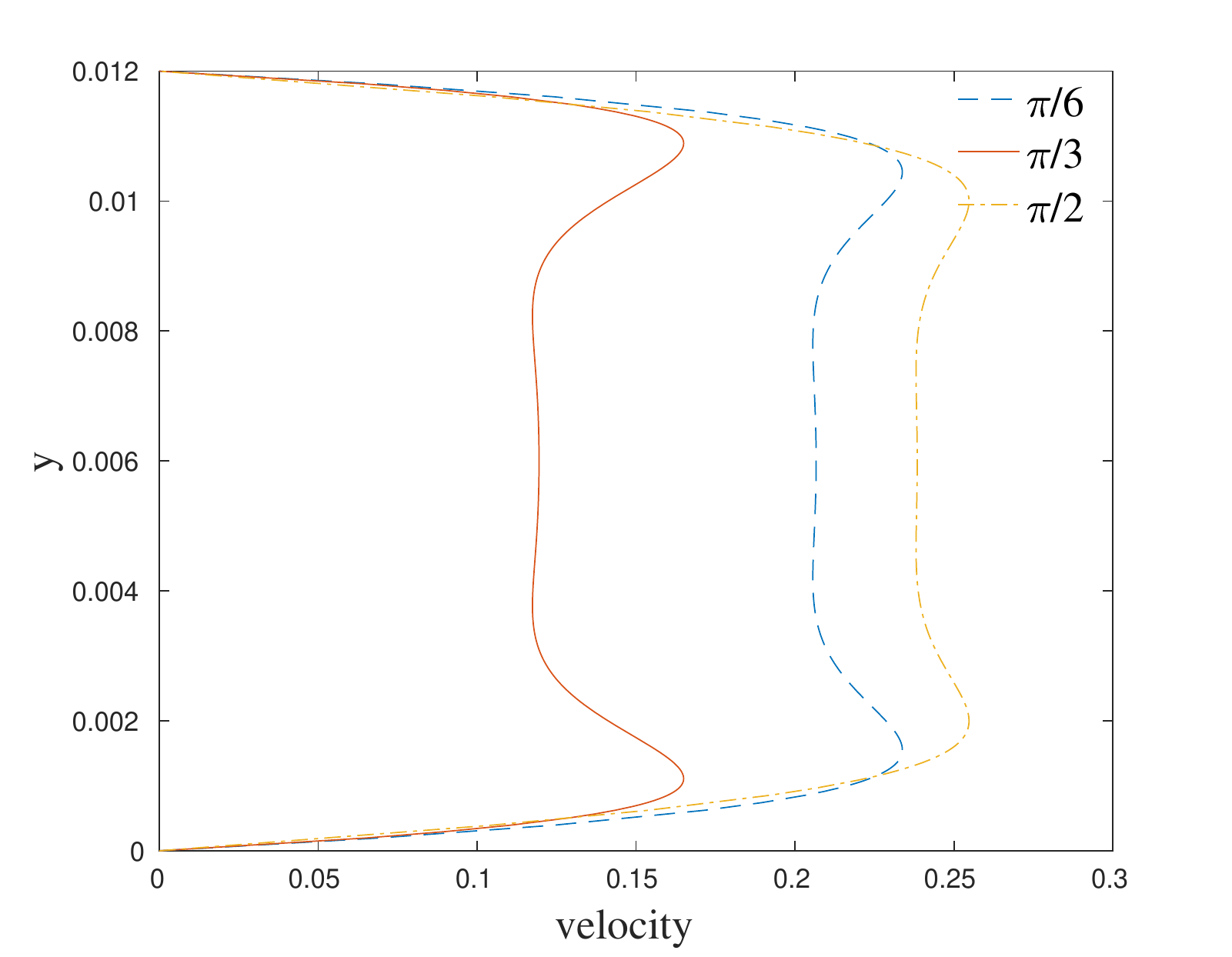}
	\caption{Velocity profiles corresponding to the computational setup used in Section \ref{subsec:valve} during a typical period:
	$Wo = 1.0$(left panel), $Wo = 5.0$(middle panel) and $Wo = 10.0$(right panel). }
	\label{figs:pulsatile-v}
\end{figure}

Considering that the pulsative flow is driven by a periodic pressure gradient 
\begin{equation*}
\frac{\partial p}{\partial x} = -A cos(\frac{2\pi}{T}t),
\end{equation*}
where $A$ denotes the pressure amplitude and $T$ the oscillation period. 
To mimic the pulsative flow in the model of Left Ventricle, we set $A_0 = 2500 N/m^3$ and $T_0 = 0.6s$.
The analytical solution for this type of flow is \cite{loudon1998use}
\begin{equation*}
\begin{split}
u_0(y,t) = \frac{A}{\omega \rho_{f} \gamma}
\bigg\{ \Big(\sinh(\Phi_1(y))  \sin(\Phi_2(y)) + \sinh(\Phi_2(y))  \sin(\Phi_1(y))  \Big) \cos(\omega t)  \\
\Big( \gamma - \cosh(\Phi_1(y))  \cos(\Phi_2(y))  - \cosh(\Phi_2(y))  \cos(\Phi_1(y)) \Big) \text{cos}(\omega t)
\bigg\},
\end{split}
\end{equation*}
where 
\begin{equation*}
\begin{cases}
\Phi_1(y)  = \frac{Wo}{\sqrt{2}} \bigg( 1 + 2.0 \frac{y - \frac{1}{2} DH}{DH} \bigg) \\
\Phi_2(y) = \frac{Wo}{\sqrt{2}} \bigg( 1 - 2.0 \frac{y - \frac{1}{2} DH}{DH} \bigg)
\end{cases},
\end{equation*}
and 
\begin{equation*}
\gamma = \cosh(\sqrt{2}W_o) + \cos(\sqrt{2}W_o).
\end{equation*}
where $W_o$ is the Womersley number. 
Fig. \ref{figs:pulsatile-v} gives velocity profiles of different $W_o$ numbers. 

\end{document}